\documentclass[aps, pra, showpacs, english, reprint, nobibnotes]{revtex4-2}
\usepackage{bbm, amsthm, bm,textcomp, nicefrac,geometry,ragged2e}
\usepackage[dvipsnames]{xcolor}
\usepackage{float}
\usepackage[bbgreekl]{mathbbol}
\usepackage{graphicx,epstopdf,color,verbatim,enumitem}
\usepackage{pbox,array}
\usepackage{mathrsfs}
\usepackage{amssymb}
\usepackage{stackrel}
\usepackage{hyperref}

\usepackage{amsmath}
\makeatletter
\usepackage[T1]{fontenc}
\usepackage{babel}
\addto\captionsenglish{}

\begin{document}

\title{Coherent randomized benchmarking}
\author{J. Miguel-Ramiro$^{1}$, A. Pirker$^{1}$ and W.~D\"ur$^1$}
\affiliation{$^1$ Institut f\"ur Theoretische Physik, Universit\"at Innsbruck, Technikerstra{\ss}e 21a, 6020 Innsbruck, Austria}

\date{\today}

\begin{abstract}
Randomized benchmarking is a powerful technique to efficiently estimate the performance and reliability of quantum gates, circuits and devices. Here we propose to perform randomized benchmarking in a coherent way, where superpositions of different random sequences rather than independent samples are used. We show that this 
leads to a uniform and simple protocol with significant advantages with respect to gates that can be benchmarked, and in terms of efficiency and scalability. We show that e.g. universal gate sets, the set of $n-$qudit Pauli operators or more general sets including arbitrary unitaries, as well as a particular $n-$qudit Clifford gate using only the Pauli set, can be efficiently benchmarked.
The price to pay is an additional complexity to add control to the involved quantum operations. However we demonstrate that this can be done by using auxiliary degrees of freedom that are naturally available in basically any physical realization, and are independent of the gates to be tested.

\end{abstract}
\maketitle

{\section{Introduction}}
With the development of quantum technologies and their widespread applications for quantum networks, computation and metrology comes the need to characterize and verify the performance of small and larger-scale quantum devices \cite{Eisert2020,kliesch2020}. This is a highly non-trivial task, as the direct classical simulation of outputs of such devices is impossible. While an exact characterization for elementary building blocks using process tomography \cite{Tomog1,Tomog2} is possible, and one can even guarantee the functionality in a device independent way \cite{Tomog3}, the required effort and complexity makes such approaches intractable for larger systems or longer quantum circuits. Randomized benchmarking (RB) \cite{Emerson2005,Levi07,Knill2008,Dankert09,Chow09,Magesan11,Magesan11_2,Kimmel2014,Epstein14,helsen2020general} is a powerful technique that allows one to estimate the average gate error of gate sets with modest overhead, where random circuits of varying length are used to extract the desired information. In this way one can separate gate errors from state preparation and measurement errors (SPAM), and access the former directly. Several extensions and modifications of RB that allow one to use fewer resources \cite{Wallman2014,Granade15,Frana2018,Roth18,Dirkse19,Boone19,Harper19,Helsen19_2,Alexander16}, or test single gates via interleaved randomized benchmarking (IRB) \cite{Magesan12,Harper2017,Onorati19,Helsen2019} have been put forward, and RB has been applied in practice \cite{Gaebler12,Corcoles13,Xia15}.
Although RB is a widely used and advantageous technique, its applicability is strongly restricted to certain sets of quantum gates (typically Clifford operations). Several works \cite{Onorati19,Helsen2019,Gambetta2012,Dugas15,Cross2016,Hashagen2018,Brown18,Erhard2019} have proposed different RB variants to overcome this problem. However these approaches are still limited in terms of the kind of gates or gate sets that can be tested, the efficiency or the scalability for systems with larger number of qubits.

Here we propose to do RB in a coherent way. Rather than sequentially testing different random sequences of gates and circuits of different length, we use a coherent superposition of them. In this way, additional averaging between different branches takes places, and the available coherence allows one to access extra information as compared to individual runs. This enables one to directly benchmark a very large variety of gates or gate-sets in a simple and uniform way, overcoming one of the main issues of RB.  What is more, the approach is efficiently scalable and one can benchmark sets of multi-qudit operations. We demonstrate our approach by considering benchmarking of universal gate sets such as Toffoli and Clifford gates, the $n$-qudit Pauli operators, $n$-qudit controlled operations and other sets which includes more general operations such as e.g. the multi-qubit Mølmer--Sørensen gate. In fact, a set containing any multi-qudit unitary operation can be designed to be tested via coherent RB. We also show that interleaved randomized benchmarking (IRB) of Clifford operations can be done using only Pauli elements. Additionally, for some of the aforementioned sets where RB is possible, one can also perform IRB for any individual gate-set element.

The coherent application of gates and circuits requires additional control and overhead though, which can however be provided by utilizing independent auxiliary operations. One can make use of auxiliary degrees of freedom that are present in basically any realization of quantum information carriers. In particular, we show proof-of-principle examples based on using spatial modes for photons \cite{Zhou2011,Arajo2014,Friis2014}, and motional degrees of freedom for trapped ions \cite{SchmidtKaler2003,Barreiro2011,Friis2014}, to add control to the applied operations. Importantly, the gates are still implemented in exactly the same way as in the standard approach, i.e. independently of the control setting, and can hence be benchmarked. This control setting can be thought as a "testing-device", e.g. a pre-calibrated factory device that allows one to test different kinds of gates and gate sets.


\section{Setting}

Standard RB protocols consist in applying a sequence of certain length of quantum gates randomly chosen from some set (usually from the Clifford group) to some known initial state, followed by their inverses. In the noiseless case, the initial state is recovered. For noisy operations, the survival fidelity ---the fidelity of the final state with respect to the initial one--- is computed by averaging over several runs, and the process is repeated for sequences of different lengths. The different  average survival fidelities are fitted to a decay curve to obtain an estimation of the average gate fidelity (see Appendix C).


The RB approach we propose here
consists in performing the different sequences in an equally-weight superposition (see Fig. \ref{fig:coherentrb}), in such a way that the coherence can be exploited to largely improve the applicability of the protocol without compromising its performance. Other approaches have focused on applicability extensions \cite{Onorati19,Helsen2019,Gambetta2012,Dugas15,Cross2016,Hashagen2018,Brown18,Erhard2019}, mainly based on exploiting mathematical properties of certain sets or groups, in order to be able to isolate the multiple fitting parameters arising in the description of non-Clifford-type groups. In contrast, our coherent RB approach retains the simplicity of the standard description with a single fitting parameter, and provides a larger applicability freedom. It can also be scaled in the number of qubits or dimension of the systems efficiently. The superpositions are achieved
following the spirit of \cite{GenuineQN} (see also \cite{Chiribella2013,Araujo2014,Procopio2015}), using  external control devices such that the unitary gates can be benchmarked independently of the control. In addition, the number of experimental runs is significantly reduced.

\subsection{Coherent RB}

\begin{figure}
\includegraphics[scale=0.5]{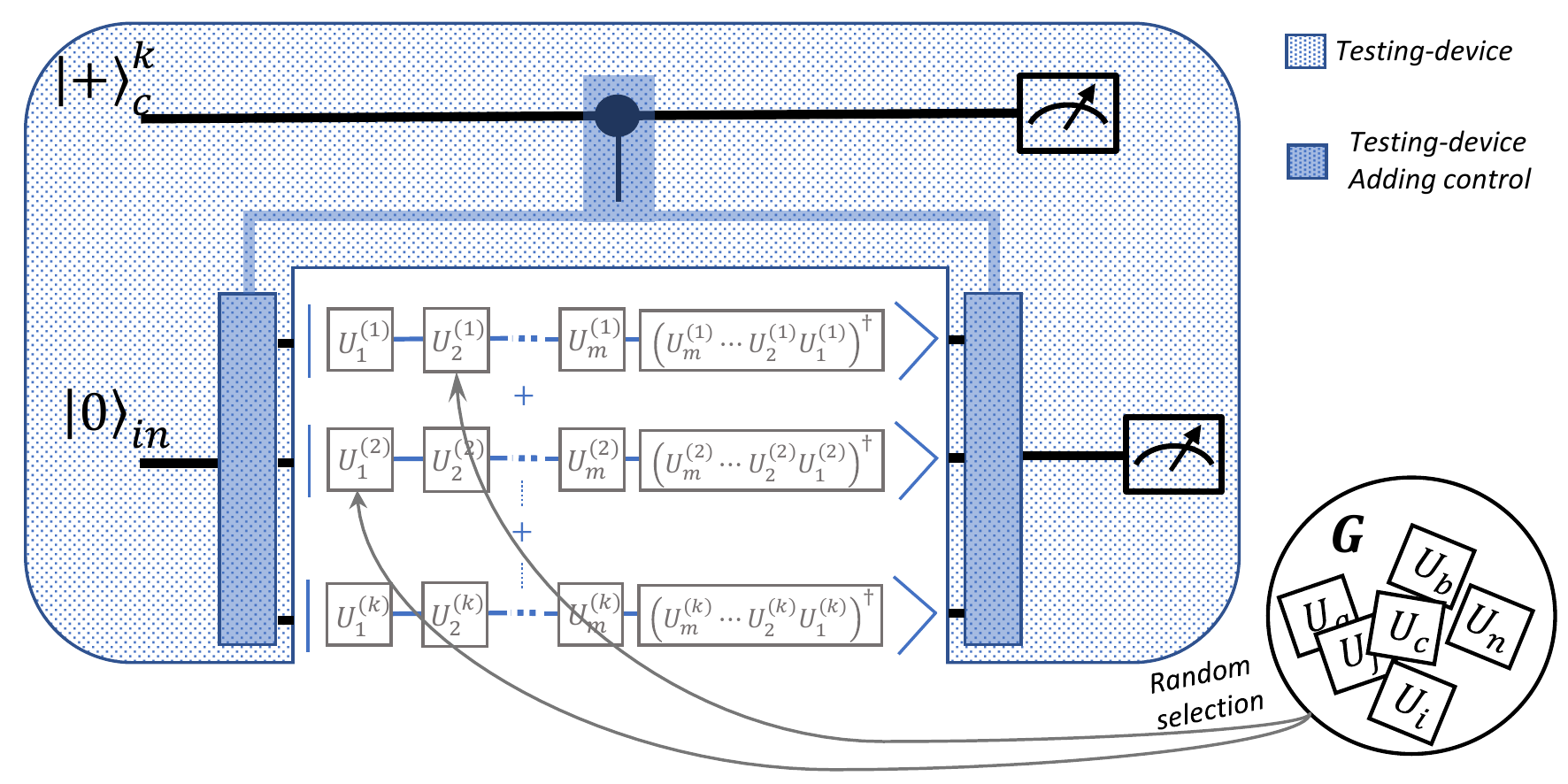}
\caption{ \label{fig:coherentrb} Coherent RB. A known initial state is evolved, controlled by a $k-$level system, into an equally-weighted superposition of $k$ different sequences of $m$ random unitary gates acting on it, i.e. $\frac{1}{\sqrt{k}}\sum_{i=1}^{k}\left|i\right\rangle _{c}\left(U_{1}^{(i)}\cdots U_{m}^{(i)}\right)\left|0\right\rangle _{in}$, where control is added with external devices. The number of experimental rounds to determine the average gate sequence fidelity is reduced by a factor of $k$. Given the extra information coming from the coherence, applicability is remarkably enhanced.}
\end{figure}

In the following we discuss the main steps of the protocol. 
Observe that, although the mathematical description is based on controlled operations, they do not have to be explicitly performed (see below) in a practical setting, since control can be added with external devices that we denote as "testing-device" (see Fig. \ref{fig:coherentrb}), while gates to be tested are placed and applied separately, as in standard RB.

\subsubsection{Protocol for coherent RB}
Consider a set $G$ of unitary operations
$U_{i}\in G$ which should be benchmarked. The
protocol proceeds as follows (see Appendix Sec. C for details).
\begin{enumerate}[label=(\roman*), leftmargin=\parindent,align=left,labelwidth=\parindent,labelsep=2pt]
\itemsep0em
\item A $k-$dimensional control register is initialized in the state $\left|+\right\rangle _{c}^{k}=\frac{1}{\sqrt{k}}\sum_{i=0}^{k-1}\left|i\right\rangle _{c}$
for some $k$, whereas the main register is initialized in
some known state, e.g. $\left|0\right\rangle $ (up to preparation errors), such that the initial state simply reads $\rho_{0}=\left|+\right\rangle _{c}^{k} \otimes \left|0\right\rangle$.

\item For some length $m$, a sequence of $m$ controlled operations
of the form $
CU^{(r)}=\sum_{i=0}^{k-1}\left|i\right\rangle _{c}\left\langle i\right|\otimes U_{i}^{(r)}$ is applied,
where $r=\left\{ 1,...,m\right\} $ defines the sequence position. Each operation $\hat{CU}^{(r)}$, where $\hat{CU}(\sigma)=CU\sigma CU^{\dagger}$ has some noise associated that we assume to be independent of the position and the superposition branch (zeroth order approximation), i.e.
\begin{equation}
S_{G}(m)=\bigcirc_{r=1}^{m}\left[{\hat{\xi}}\circ\hat{CU}^{(r)}\right]\left(\rho_{0}\right),\label{eq:CUseq}
\end{equation}
where we restrict for the moment to the case of noiseless control register, i.e. $\hat{\xi}(\rho)=I_{d}\otimes\sum_{i,j}\chi_{ij}\mathcal{\mathrm{P}}_{i}\rho\mathcal{\mathrm{P}}_{j}^{\dagger}$, where $\chi$ is the channel matrix written in the $n-$qudit Pauli basis  (see Appendix A). 
Following the "testing-device" reasoning, the ideal implementation of the control register part is well justified, since control can be added by external devices (see below) which can be already well-calibrated. Noise in the control register can however be included in the protocol description, without altering its simplicity or performance (see below).

Importantly, each $n-$qudit unitary $U_{i}^{(r)}$ acting at each branch and position is taken randomly from a uniform probability distribution of the set $G$. Observe that a coherent superposition of
$k$ random sequences of gates is achieved.

\item For each branch $i$ of the superposition, the inverse of the branch sequence
is computed, i.e. $U_{i}^{(m+1)}=\left(U_{i}^{(m)}\cdots U_{i}^{(1)}\right)^{\dagger}$ and applied at the $(m+1)_{th}$ position. This is achieved by
a controlled operation $CU^{(m+1)}=\sum_{i=0}^{k-1}\left|i\right\rangle _{c}\left\langle i\right|\otimes U_{i}^{(m+1)}$. In the absence of noise, the final state would be mapped back to the initial one $\left|+\right\rangle _{c}^{k} \otimes \left|0\right\rangle$. Assuming that this $(m+1)^{\rm th}$-controlled gate introduces some noise $\hat{\xi}^{(j+1)}$ independent of the sequence, the final state reads
\begin{equation}
\rho_{f}=\left(\hat{I}\otimes {\hat{\xi}}^{(m+1)}\right)\circ\hat{CU}^{(m+1)}\circ\left(\bigcirc_{j=1}^{m}\left[ {\hat{\xi}}\circ\hat{CU}^{(j)}\right]\left(\rho_{0}\right)\right).\label{eq:finalstate}
\end{equation}

\item A binary-outcome POVM $\left\{ E_{\psi},{I}-E_{\psi}\right\}$ is performed to measure the final state taking into account measurement errors, such that in the ideal-measurement case the POVM equals a projective measurement $\left\{ \left|\psi\right\rangle \left\langle \psi\right|,{I}-\left|\psi\right\rangle \left\langle \psi\right|\right\} $,
where $\left|\psi\right\rangle =\left|+\right\rangle _{c}^{k}\otimes\left|0\right\rangle _{in}$.
The result of the measurement gives us an estimation of the average sequence coherent fidelity, i.e.
\begin{equation}
F_{G}\left(m,k\right) = \mathrm{tr}\left(E_{\psi}\rho_{f}\right),\label{eq:fidelity1}
\end{equation}
so that  $F_{G}\left(m,k\right) \approx F_{G}\left(m\right)$, where $F_{G}\left(m\right)$ is the exact average sequence fidelity obtained when considering a superposition of all the different $k=|G|^{m}$ sequences.

\item The process is repeated for different values of $m$ such that the average sequence coherent fidelities approach the decay curve (see Appendix C)
\begin{equation}
F_{G}\left(m\right)=A\chi_{00}^{m},\label{eq:finalfidelity}
\end{equation}
with $A=\mathrm{tr}\left(E_{\psi}\left(\hat{I}\otimes {\xi}^{(j+1)}\right)(\rho_{0})\right)$
a constant independent of $m$ that absorbs SPAM errors. Ideally, $A=1.$ Note that the parameter $\chi_{00}$ is directly related to the average gate fidelity of the set $G$ \cite{Carignan_Dugas_2019}. As shown in Appendix C, Eq. (\ref{eq:finalfidelity}) fits for any set of gates $U_{i} \in G$
of size $|G|$  which fulfills the condition
\begin{equation}
\sum_{i=1}^{|G|}U_{i}^{\dagger}P_{\mathbf{j}}U_{i}=\begin{cases}
|G|I_{d}^{\otimes n} & \mathbf{j}=\mathbf{o}\\
0 & \mathbf{j}\neq\mathbf{o}
\end{cases},\label{eq: condition}
\end{equation}
for any Pauli element
$P_{\mathbf{j}}\equiv X^{\mathbf{j}_{r}}Z^{\mathbf{j}_{s}}=X^{s_{1}}Z^{r_{1}}\otimes\cdots\otimes X^{s_{n}}Z^{r_{n}}$ and with $P_{\mathbf{o}}=I^{\otimes n}$. Observe that the Pauli elements come from the noisy channel representation, and therefore can be replaced by any $n-$qudit basis operators.
\end{enumerate}

\subsubsection{Gate sets that can be benchmarked}
Condition Eq. (\ref{eq: condition}) is easy to check, and allows us to show that the coherent approach offers remarkably broader freedom to benchmark different sets of quantum gates as compared to previous approaches, and keeping the simplicity of the description. Some instances of sets that fit  Eq. (\ref{eq: condition}) and can be benchmarked via coherent RB are (see Appendix B):
\begin{enumerate}[label=$\circ$]
\itemsep0em
\item The $n-$qudit Clifford group.
\item The set of $n-$qudit Pauli operators.
\item A set of $n-$qudit controlled operations.
\item Multi-qubit Mølmer--Sørensen type gate sets.
\item Any $n-$qudit unitary set constructed as $P_{i} U$.
\end{enumerate}
In contrast to standard approaches, where twirling can be seen as a group
averaging, coherent RB can be understood as a "multiple" gate-set averaging
which is performed in a single round due to the extra power of the generated coherence. The price to pay for adding control to operations (see below) is an overhead complexity which scales linearly with the number of qudits $n$, and with the number of sequences in superposition $k$ (or dimension of the control register). Note however that this linear overhead with $k$ also appears in standard RB as the number of independent sequences over which the average is performed \cite{Wallman2014,Helsen19_2}.

\subsubsection{Further advantages of coherent RB. Efficiency and scalability}
Since generating a superposition comprising all $|G|^{m}$ sequences is clearly inefficient, one considers a certain number $k\ll|G|^{m}$ of sequences that leads to an estimated average fidelity with some deviation confidence. This confidence region of size $\epsilon$
arises, such that the probability that the estimated fidelity lies
within this confidence region is greater than some set confidence
level $1-\delta$, i.e. $P\left[\left|F_{k}-\bar{F}\right|<\epsilon\right]\geq1-\delta$. Although we leave a detailed statistical analysis for future work, numerical analyses  indicate that this confidence region, and therefore the efficiency of the coherent approach, i.e. the value of $k$ for a fixed confidence, can be significantly better in some regimes of interest, when compared with standard approaches. Moreover, when applying coherent RB to gate-sets outside the Clifford group, the magnitude orders of the estimated fidelity show that the protocol efficiency is not compromised for these sets, that cannot be benchmarked with standard RB protocols. We refer to Appendix E for details.

All these results also apply for systems of increased dimension and number, i.e. for testing $n-$qudit quantum gates, for which the efficiency of the protocol is not compromised. This implies another important difference with other RB approaches, which are in general inefficient when the tested gates involve higher dimensional quantum systems or multi-systems.

There is another remarkable advantage of coherent RB, namely a reduction of the number of required rounds of the experiment as compared to standard RB. By performing coherent RB in superposition of $k$ different instances, the reduction factor is given by $k$. Therefore, following the protocol description, it is direct to see that the total number of measurements that need to be performed is significantly lower in our coherent approach, because of the reduction of the number of runs or repetitions needed for each sequence length $m$ experiment. For settings where measurements are slow or costly, this yields a significant advantage.

\subsection{Interleaved coherent RB}

A RB variant of particular interest is interleaved randomized benchmarking (IRB). It was developed \cite{Harper2017,Magesan12,Onorati19,Helsen2019} in
order to benchmark particular
quantum gates. This is achieved by interleaving random operations from a set and
the particular gate, and comparing the average sequence fidelity
to the case without the interleaved operation. An immediate application of our coherent RB approach is that one can benchmark via coherent IRB quantum gates from certain sets mentioned above, such as  e.g. $n-$qudit Pauli gates. In addition, the required set size can be reduced, which we demonstrate by showing coherent IRB of a $n-$qudit Clifford gate using the set of Pauli operations only, an interesting application of our coherent approach that again is not possible with standard IRB.

\subsubsection{Interleaved coherent RB of Clifford gates using Pauli operations}
Consider an arbitrary Clifford gate $C\subseteq\mathrm{\mathit{\mathcal{C}}}_{n}$, and consider the $n-$qudit Pauli gates $P_{\mathbf{i}}\in\mathcal{P}_{d,n}$. In order to benchmark the Clifford gate $C$ we perform the following
steps:
\begin{enumerate}[label=(\roman*), leftmargin=\parindent,align=left,labelwidth=\parindent,labelsep=2pt]
\item The average  gate coherent fidelity of the Pauli set is computed as
the reference fidelity. This is accomplished following exactly the steps
of coherent RB explained above, such that the parameter $\chi_{00}^{P}$ is
found.
\item The same process is repeated, but with the quantum gate $C$
interleaved every second position in all the branches of the superposition. This is achieved by the operation
$\hat{CU}_{C}=\hat{I}\otimes \hat{C}=\hat{U}_{C}.$ For some known initial state $\rho_{0}=\left|+\right\rangle _{c}^{d}\left\langle +\right|\otimes\left|0\right\rangle _{in}\left\langle 0\right|$
(up to preparation errors), the system is evolved to
\begin{equation}
\rho_{f}=\hat{CU}^{(m+1)}\circ\left(\bigcirc_{j=1}^{m}\left[ {\hat{\xi}_{C}}\circ\hat{U}_{C}\circ {\hat{\xi}}\circ\hat{CU}^{(j)}\right]\left(\rho_{0}\right)\right),\label{eq:interv2}
\end{equation}
where now $\hat{CU}^{(m+1)}$ is the inverse
including the interleaved ${U}_{C}$ operation. We assume that this $(m+1)^{\rm th}$ gate is noiseless for simplicity. Given the fact that $C^{\dagger}P_{i}C$ is mapped to another element of $\mathcal{P}_{d,n}\setminus I^{\otimes n}$, the noise matrix $\chi^{C}$ associated to the gate $C$ is mapped to another matrix $\chi^{\bar{C}}$, with the element $\chi_{00}$ invariant. Thus it can be shown (see Appendix D) that the average sequence interleaved coherent fidelity reads
\begin{equation}
F_{P,\bar{C}}\left(m\right)=\left(\chi_{00}^{P\circ\bar{C}}\right)^{m}.\label{eq:Fpc3}
\end{equation} \end{enumerate}
One can obtain an estimation of the desired
parameter $\chi_{00}^{C}$ noting that $\chi_{00}^{C}=\chi_{00}^{\bar{C}}$
and that \cite{Kimmel2014} $\chi_{00}^{P\circ\bar{C}}=\chi_{00}^{P}\chi_{00}^{\bar{C}}\pm E$,
with some estimation bound $E$ (see Appendix D for details). This bound is particularly tight in the regime of interest where Pauli gates have much larger fidelity than the
$C$ operation, and therefore  we can efficiently benchmark a Clifford gate using just Pauli operators, highlighting the applicability flexibility of coherent RB.

\subsubsection{Alternative method for arbitrary gates}
Observe that an alternative approach for benchmarking a particular arbitrary gate is possible. Via standard coherent RB, we can benchmark the set of operations $M_{\mathbf{i}}=P_{\mathbf{i}}U$ (see Appendix B), for any (set of) $n-$qudit unitary operation(s) $U$ and all the $n-$qudit Pauli elements $P_{\mathbf{i}}$. Under the assumption of perfect implementation of Pauli gates, one can, by standard coherent RB, directly access the noise of the gate(s) $U$. This assumption is valid in many physical set-ups, as such single-qudit gates are often much easier to implement than more general operations as e.g. entangling gates.


\subsection{Noise in the control register}

Although, given the testing-device setting stressed above one can assume high reliability on the part
of the setting responsible for adding control, in such a way that
the assumption of noiseless control register in coherent RB
can be justified, we show how the process can be also insensitive to this noise. Consider that, after each $CU^{(j)}$ operation, the control register
is affected by a depolarizing noise with parameter $q$, of the form
$\zeta\left(\rho\right)=q\rho+\frac{1-q}{k}I,$
where $k$ is the dimension of the control register or, equivalently,
the number of sequences in the superposition.
The average sequence coherent fidelity for a fixed length $m$ thus becomes
\begin{equation}
F_{G}\left(m\right)\approx\left(q\cdot\chi_{00}\right)^{m}+\frac{\left(1-q^{m}\right)}{k} f_{G},\label{eq:fideerrorcontrol}
\end{equation}
where we ignore preparation and measurement errors for simplicity, and where $f_{G}$ is the average state fidelity of a set, defined as  $f_{G}=\left\langle \left| \left\langle\varphi\right| \mathcal{U}\left|\varphi\right\rangle \right| ^ {2} \right\rangle $, with the average over all the (noiseless) unitary operations of the set $G$ with respect to the state $\left|\varphi\right\rangle$. For instance, for the set of $n-$qudit Pauli operators it is direct to see that $f_{G}=\frac{1}{d^{n}}$, with $d$ and $n$ the dimension and number of target qudits respectively.

This can be understood as follows. The left part of Eq. (\ref{eq:fideerrorcontrol})
corresponds to the ``correct'' branch of the process, i.e. with
probability $q^{m}$ the output equals the noiseless-control protocol. The right part
of the expression represents the ``error'' branch of the process, which
occurs with probability $1-q^{m}$. After each $CU^{(j)}$
operation, within this error branch, exponentially  (with $m$) many sub-branches of random sequences are found, which ultimately do not correspond
to the inverse operations applied in the $(m+1)^{th}$ position. The final states associated to each sub-branch
correspond to random uniformly distributed operations applied from the set $G$, to
the initial state, with an average fidelity $f_{G}$, whose estimation deviation influence can be assumed negligible with respect to the protocol deviation.  The factor $\frac{1}{k}$ comes from the fact
that only incoherent terms contribute to the right term of Eq. (\ref{eq:fideerrorcontrol}).

Observe that, although the system register is not directly affected by noise in our model, the noise actually spreads and leads to correlated noise, since each of the “local” errors affects the other register in the next sequence position (and in all later operations that follow) within the same protocol.

Noise in the control register can be hence included in the protocol
description, and the process can still be realized under the assumption
that the control noise parameter $q$ is known with some known accuracy. This is justified from the proof-of-concept physical implementations, where the control
setting (testing-device) can be first tested by applying controlled-identities, i.e. only the operations that add control.


\section{Adding control with external devices}

An important feature of coherent RB is that control can be added with external devices from a practical point of view, such that the quantum gates to be benchmarked are performed in exactly the same way as in the standard RB approach. The device setting responsible for adding control (also for state preparation and measurements) can conform a previously well calibrated testing-device, that can be used to test many different gates or gate-sets for e.g. factory applications where the quality of fabricated devices should be tested. We show here proof-of-principle examples of how control can be added with external devices. Despite the current technology limitations, we believe that this approach can be very useful in the mid and long timescale, possible using optimized experimental settings based on e.g. non-linear optics \cite{Lin2015} or superconducting qubits \cite{Friis2015S}.

\subsection{Proof of principle implementation with linear optics}

\begin{figure*}
 \centering
    \includegraphics[width=\linewidth]{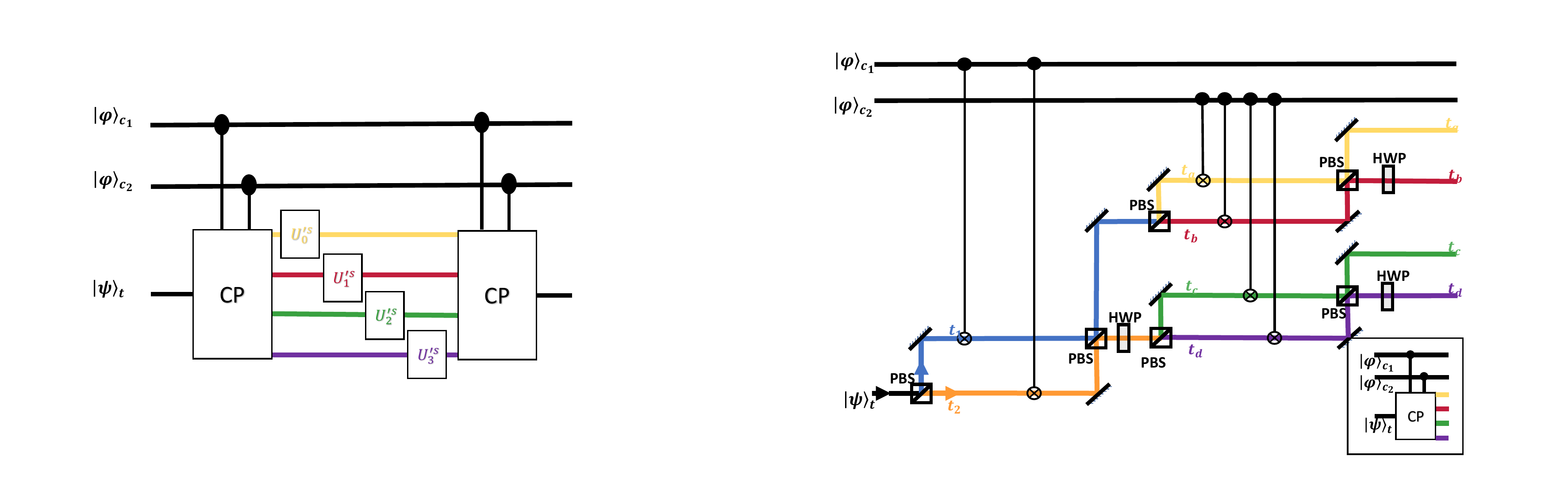}

\caption{\label{fig:opticalim2} Left figure shows the general optical setting for achieving the desired superposition of four sequences with external device control. Right figure details the control-path (CP) gate of the left setting.}
\end{figure*}

The first proof-of-principle argument follows from \cite{Zhou2011,Arajo2014,Friis2014,rubino2020} and is based on a linear optics implementation. $d$
control photons are initially prepared in  the polarization state $\left(\left|H\right\rangle+\left|V\right\rangle\right)^{\otimes {d}}$, playing the joint role of an effective $k=2^{d}$ dimensional control
register. The target photons are also prepared in the polarization state. A controlled-path (CP) gate, a generalization of \cite{Zhou2011}, is applied from the control qubits. Each CP gate makes the target photon pass through several photon beam splitters (PBS), followed by CNOT
gates which are applied interspersed from the control, and final
PBSs and half-waveplates that adequately mix and separate the spatial modes of the target depending on the joint control state.
The gate sequences of the operations to be tested are applied independently on the different spatial modes of the target. By undoing the CP gate,  the superposition is achieved, and crucially,
with external-device control only.  Fig. \ref{fig:opticalim2}
shows an example where an equally-weight superposition of four different
sequences of unitary operations from a given set is achieved.  The target photon is prepared in some polarization state $\left|\psi\right\rangle _{t}=\alpha\left|H\right\rangle _{t}+\beta\left|V\right\rangle _{t}.$
The initial state reads $\left|\varphi\right\rangle _{c}\left|\psi\right\rangle _{t}$. A controlled-path (CP) gate is applied from the
two control photonic qubits.  The state after the CP gates reads
\small
\begin{align}
&\frac{1}{2}\left|H\right\rangle _{c_{1}}\left|H\right\rangle _{c_{2}}\left|\psi\right\rangle _{t_{a}}+\frac{1}{2}\left|H\right\rangle _{c_{1}}\left|V\right\rangle _{c_{2}}\left|\psi\right\rangle _{t_{b}}+ \nonumber \\
&\frac{1}{2}\left|V\right\rangle _{c_{1}}\left|H\right\rangle _{c_{2}}\left|\psi\right\rangle _{t_{c}}+\frac{1}{2}\left|V\right\rangle _{c_{1}}\left|V\right\rangle _{c_{2}}\left|\psi\right\rangle _{t_{d}},
\label{eq:CP}
\end{align}\normalsize
Subsequently, the four different sequences
of length $m$ of randomly chosen unitary operations $\left(U_{i}^{(j)}\right)$ are applied independently on  the four beam branches,
each sequence affecting only one spatial mode. By finally undoing
the CP operations, one finds the desired superposition
\small
\begin{align}
\frac{1}{2} \sum_{i=0}^{3}\left|i\right\rangle _{c}\left(U_{i}^{(1)}\cdots U_{i}^{(m)}\right)\left|\psi\right\rangle _{t}.
\end{align}\normalsize
Although the apparent restriction to power-of-two dimensions, any
effective value of $k$ can be reached by just avoid applying any
unitary on some of the branches (or equivalently applying identities)
and taking it into account when studying the
sequence fidelity. Observe also that the dimension of the
Hilbert space associated to the spatial modes of the photons can be made arbitrarily large. Note that the complexity, in terms of the number
of logic gates required, scales linearly with the dimension of the
control register $k$ and linearly with the
number of target qubits.

\subsection{Proof of principle implementation with trapped ions}

A similar scheme works for trapped ions, where motional degrees of freedom, or other auxiliary energy levels, can be used to implement the superposition. The random gates to be tested are implemented by separate laser pulses. We make use of the fact that sideband pulses, as well as hiding pulses that move the logical qubit space to some auxiliary space, are available (see \cite{Friis2014} for a similar setting). We consider $n+m$ ions, each with four internal energy levels, $\{|g\rangle,|e\rangle\}$ that form the logical qubit space, and two additional levels $\{|g'\rangle,|e'\rangle\}$ with different energy spacing, such that the ions share a common vibrational mode. The first $n$ ions $A_1,A_2, \ldots A_n$ encode the $2^n$ superposition branches, and the remaining $m$ ions $B_1,B_2, \ldots B_m$ the system to be tested, which is initially in the state $|\psi\rangle$. The superposition is transferred from the first to the second ones via some hiding pulses $S$ that do not change the motional degree of freedom. Blue and red detuned pulses are used to selectively increase the motional excitation depending on the internal state of the ion as in the Cirac-Zoller gate \cite{CiracZoller}, acting as identity on remaining levels, so that the sequences of controlled gates can be applied with external control setting. By undoing the steps, the desired superposition is obtained.

We provide an example to illustrate our scheme for $n=2$ and $m=1$, i.e. a superposition of four branches and gates acting on a single qubit. We make use only of standard control techniques that are available in different ion-trap set-ups, and have been demonstrated and utilized in other contexts \cite{SchmidtKaler2003,Barreiro2011}.

We consider hiding pulses $S_g=(|g\rangle\langle g'| + |g'\rangle\langle g|) \otimes I$ and $S_e=(|e\rangle\langle e'| + |e'\rangle\langle e|) \otimes I$ which do not change the motional degree of freedom. We use blue detuned hiding pulses $H_{b}^j$ that map $|g\rangle\otimes|k\rangle \leftrightarrow |g'\rangle\otimes |k-j\rangle$ and $|e\rangle\otimes|k\rangle \leftrightarrow |e'\rangle\otimes |k-j\rangle$ (note that if $k-j<0$, no transition takes place since there is no corresponding energy level; $j$ denotes the $j^{\rm th}$ sideband), and red detuned pulses $H_{r}^j$ that map $|g\rangle\otimes|k\rangle \leftrightarrow |g'\rangle\otimes |k+j\rangle$ and $|e\rangle\otimes|k\rangle \leftrightarrow |e'\rangle\otimes |k+j\rangle$. In addition we consider red detuned pulses that selectively increase the motional excitation depending on the internal state of the ion, $S_g^j$, which maps $|g\rangle\otimes|k\rangle \leftrightarrow |e\rangle\otimes|k+j\rangle$ and acts as identity on remaining levels.

We start by initializing the system in $|0_x\rangle_{A_1}|0_x\rangle_{A_2}|\psi\rangle_{B_1}\otimes|0\rangle$, where $|0_x\rangle=(|e\rangle +|g\rangle)/\sqrt{2}$. By using $S_{gA_1}^1$ and $S_{gA_2}^2$, we transfer the superposition to the motional degree of freedom, resulting into $|e\rangle_{A_1}|e\rangle_{A_2}|\psi\rangle\otimes (|0\rangle +|1\rangle +|2\rangle + |3\rangle)/2$. To apply now a gate, or a circuit, in one particular branch $k$, we proceed as follows. First we use $H_{b}^{k+1}$ to hide all states with a motional degree of freedom $l>k$, and then $H_r^{n-k-1}$ followed by $H_b^n$. This results into only the selected state with initial motional degree $|k\rangle$ to be transferred to $|\psi\rangle_{B_1} \otimes|0\rangle$, while for all other motional degrees $j$, the state is transferred to the hiding subspace $\{|g'\rangle,|e'\rangle\}_{B_1}$. One can now apply the gate(s) corresponding to the $k^{\rm th}$ branch on the system, and then undo above steps. The different branches are then considered sequentially, using above method for $k=0,1 \ldots 2^n-1$, such that the desired superposition is achieved.

Observe also that, as pointed out in \cite{Friis2014}, devices or laser beams that implement the unitary gates can be reused to apply the operations at different positions and branches. This can be done in such a way that each operation, even if it appears in multiple branches of the superposition, is performed only once, leading to further reduction of complexity. We remark that these are proof-of-concept arguments that justify the feasibility of adding control with external devices in a practical setting, where however further optimization is required for an experimental implementation.



{\section{Conclusions}
We have introduced an alternative approach to randomized benchmarking (RB) of quantum gates, i.e. coherent RB, which consists in realizing the protocol in superposition of the different sequences. The superposition is achieved by adding control to the operations. Coherent RB leads to several remarkable advantages, both in the standard and interleaved variants. Our coherent approach is largely more flexible, allowing to benchmark gate sets that can otherwise not be tested with standard RB without compromising the scalability and efficiency of the protocol. Some relevant examples that can be benchmarked are the set of  $n-$qudit Pauli operators, $n-$qudit controlled operations or more general sets including arbitrary unitaries.  Coherent RB retains the simplicity of the standard description and can be scaled in the number of qubits or dimension of the systems without compromising the efficiency.  In addition, when compared with standard approaches, coherent RB can significantly enhance the efficiency in certain regimes, while also reducing the number of experimental runs by a factor $k$, since all different sequences are processed in a single run and not sequentially.

Finally, we provide proof-of-concept arguments which show that one can add control with external devices (testing-device) that can either be much better controlled or independently benchmarked, such that gates to be tested are implemented as in the standard case.
We believe that this proposal to use coherent control in classical testing strategies is widely applicable also beyond RB, and may open a new avenue to design more efficient and reliable methods to verify and validate devices, measure state and gate fidelities or analyze the performance of quantum channels and quantum networks.


\section*{Acknowledgements} We thank B. Kraus for interesting discussions. This work was supported by the Austrian Science Fund (FWF) through project P30937-N27.

\bibliographystyle{apsrev4-2}
\bibliography{RBBiblio}

\begin{thebibliography}{55}%
\makeatletter
\providecommand \@ifxundefined [1]{%
 \@ifx{#1\undefined}
}%
\providecommand \@ifnum [1]{%
 \ifnum #1\expandafter \@firstoftwo
 \else \expandafter \@secondoftwo
 \fi
}%
\providecommand \@ifx [1]{%
 \ifx #1\expandafter \@firstoftwo
 \else \expandafter \@secondoftwo
 \fi
}%
\providecommand \natexlab [1]{#1}%
\providecommand \enquote  [1]{``#1''}%
\providecommand \bibnamefont  [1]{#1}%
\providecommand \bibfnamefont [1]{#1}%
\providecommand \citenamefont [1]{#1}%
\providecommand \href@noop [0]{\@secondoftwo}%
\providecommand \href [0]{\begingroup \@sanitize@url \@href}%
\providecommand \@href[1]{\@@startlink{#1}\@@href}%
\providecommand \@@href[1]{\endgroup#1\@@endlink}%
\providecommand \@sanitize@url [0]{\catcode `\\12\catcode `\$12\catcode
  `\&12\catcode `\#12\catcode `\^12\catcode `\_12\catcode `\%12\relax}%
\providecommand \@@startlink[1]{}%
\providecommand \@@endlink[0]{}%
\providecommand \url  [0]{\begingroup\@sanitize@url \@url }%
\providecommand \@url [1]{\endgroup\@href {#1}{\urlprefix }}%
\providecommand \urlprefix  [0]{URL }%
\providecommand \Eprint [0]{\href }%
\providecommand \doibase [0]{https://doi.org/}%
\providecommand \selectlanguage [0]{\@gobble}%
\providecommand \bibinfo  [0]{\@secondoftwo}%
\providecommand \bibfield  [0]{\@secondoftwo}%
\providecommand \translation [1]{[#1]}%
\providecommand \BibitemOpen [0]{}%
\providecommand \bibitemStop [0]{}%
\providecommand \bibitemNoStop [0]{.\EOS\space}%
\providecommand \EOS [0]{\spacefactor3000\relax}%
\providecommand \BibitemShut  [1]{\csname bibitem#1\endcsname}%
\let\auto@bib@innerbib\@empty
\bibitem [{\citenamefont {Eisert}\ \emph {et~al.}(2020)\citenamefont {Eisert},
  \citenamefont {Hangleiter}, \citenamefont {Walk}, \citenamefont {Roth},
  \citenamefont {Markham}, \citenamefont {Parekh}, \citenamefont {Chabaud},\
  and\ \citenamefont {Kashefi}}]{Eisert2020}%
  \BibitemOpen
  \bibfield  {author} {\bibinfo {author} {\bibfnamefont {J.}~\bibnamefont
  {Eisert}}, \bibinfo {author} {\bibfnamefont {D.}~\bibnamefont {Hangleiter}},
  \bibinfo {author} {\bibfnamefont {N.}~\bibnamefont {Walk}}, \bibinfo {author}
  {\bibfnamefont {I.}~\bibnamefont {Roth}}, \bibinfo {author} {\bibfnamefont
  {D.}~\bibnamefont {Markham}}, \bibinfo {author} {\bibfnamefont
  {R.}~\bibnamefont {Parekh}}, \bibinfo {author} {\bibfnamefont
  {U.}~\bibnamefont {Chabaud}},\ and\ \bibinfo {author} {\bibfnamefont
  {E.}~\bibnamefont {Kashefi}},\ }\href
  {https://doi.org/10.1038/s42254-020-0186-4} {\bibfield  {journal} {\bibinfo
  {journal} {Nat. Rev. Phys.}\ }\textbf {\bibinfo {volume} {2}},\ \bibinfo
  {pages} {382} (\bibinfo {year} {2020})}\BibitemShut {NoStop}%
\bibitem [{\citenamefont {Kliesch}\ and\ \citenamefont
  {Roth}(2020)}]{kliesch2020}%
  \BibitemOpen
  \bibfield  {author} {\bibinfo {author} {\bibfnamefont {M.}~\bibnamefont
  {Kliesch}}\ and\ \bibinfo {author} {\bibfnamefont {I.}~\bibnamefont {Roth}},\
  }\href@noop {} {\bibfield  {journal} {\bibinfo  {journal} {e-print arXiv:
  2010.05925 [quant-ph]}\ } (\bibinfo {year} {2020})}\BibitemShut {NoStop}%
\bibitem [{\citenamefont {Poyatos}\ \emph {et~al.}(1997)\citenamefont
  {Poyatos}, \citenamefont {Cirac},\ and\ \citenamefont {Zoller}}]{Tomog1}%
  \BibitemOpen
  \bibfield  {author} {\bibinfo {author} {\bibfnamefont {J.~F.}\ \bibnamefont
  {Poyatos}}, \bibinfo {author} {\bibfnamefont {J.~I.}\ \bibnamefont {Cirac}},\
  and\ \bibinfo {author} {\bibfnamefont {P.}~\bibnamefont {Zoller}},\ }\href
  {https://doi.org/10.1103/PhysRevLett.78.390} {\bibfield  {journal} {\bibinfo
  {journal} {Phys. Rev. Lett.}\ }\textbf {\bibinfo {volume} {78}},\ \bibinfo
  {pages} {390} (\bibinfo {year} {1997})}\BibitemShut {NoStop}%
\bibitem [{\citenamefont {Mohseni}\ \emph {et~al.}(2008)\citenamefont
  {Mohseni}, \citenamefont {Rezakhani},\ and\ \citenamefont {Lidar}}]{Tomog2}%
  \BibitemOpen
  \bibfield  {author} {\bibinfo {author} {\bibfnamefont {M.}~\bibnamefont
  {Mohseni}}, \bibinfo {author} {\bibfnamefont {A.~T.}\ \bibnamefont
  {Rezakhani}},\ and\ \bibinfo {author} {\bibfnamefont {D.~A.}\ \bibnamefont
  {Lidar}},\ }\href {https://doi.org/10.1103/PhysRevA.77.032322} {\bibfield
  {journal} {\bibinfo  {journal} {Phys. Rev. A}\ }\textbf {\bibinfo {volume}
  {77}},\ \bibinfo {pages} {032322} (\bibinfo {year} {2008})}\BibitemShut
  {NoStop}%
\bibitem [{\citenamefont {P\'al}\ \emph {et~al.}(2014)\citenamefont {P\'al},
  \citenamefont {V\'ertesi},\ and\ \citenamefont {Navascu\'es}}]{Tomog3}%
  \BibitemOpen
  \bibfield  {author} {\bibinfo {author} {\bibfnamefont {K.~F.}\ \bibnamefont
  {P\'al}}, \bibinfo {author} {\bibfnamefont {T.}~\bibnamefont {V\'ertesi}},\
  and\ \bibinfo {author} {\bibfnamefont {M.}~\bibnamefont {Navascu\'es}},\
  }\href {https://doi.org/10.1103/PhysRevA.90.042340} {\bibfield  {journal}
  {\bibinfo  {journal} {Phys. Rev. A}\ }\textbf {\bibinfo {volume} {90}},\
  \bibinfo {pages} {042340} (\bibinfo {year} {2014})}\BibitemShut {NoStop}%
\bibitem [{\citenamefont {Emerson}\ \emph {et~al.}(2005)\citenamefont
  {Emerson}, \citenamefont {Alicki},\ and\ \citenamefont
  {{\.{Z}}yczkowski}}]{Emerson2005}%
  \BibitemOpen
  \bibfield  {author} {\bibinfo {author} {\bibfnamefont {J.}~\bibnamefont
  {Emerson}}, \bibinfo {author} {\bibfnamefont {R.}~\bibnamefont {Alicki}},\
  and\ \bibinfo {author} {\bibfnamefont {K.}~\bibnamefont {{\.{Z}}yczkowski}},\
  }\href {https://doi.org/10.1088/1464-4266/7/10/021} {\bibfield  {journal}
  {\bibinfo  {journal} {J. Opt. B}\ }\textbf {\bibinfo {volume} {7}},\ \bibinfo
  {pages} {S347} (\bibinfo {year} {2005})}\BibitemShut {NoStop}%
\bibitem [{\citenamefont {L\'evi}\ \emph {et~al.}(2007)\citenamefont {L\'evi},
  \citenamefont {L\'opez}, \citenamefont {Emerson},\ and\ \citenamefont
  {Cory}}]{Levi07}%
  \BibitemOpen
  \bibfield  {author} {\bibinfo {author} {\bibfnamefont {B.}~\bibnamefont
  {L\'evi}}, \bibinfo {author} {\bibfnamefont {C.~C.}\ \bibnamefont {L\'opez}},
  \bibinfo {author} {\bibfnamefont {J.}~\bibnamefont {Emerson}},\ and\ \bibinfo
  {author} {\bibfnamefont {D.~G.}\ \bibnamefont {Cory}},\ }\href
  {https://doi.org/10.1103/PhysRevA.75.022314} {\bibfield  {journal} {\bibinfo
  {journal} {Phys. Rev. A}\ }\textbf {\bibinfo {volume} {75}},\ \bibinfo
  {pages} {022314} (\bibinfo {year} {2007})}\BibitemShut {NoStop}%
\bibitem [{\citenamefont {Knill}\ \emph {et~al.}(2008)\citenamefont {Knill},
  \citenamefont {Leibfried}, \citenamefont {Reichle}, \citenamefont {Britton},
  \citenamefont {Blakestad}, \citenamefont {Jost}, \citenamefont {Langer},
  \citenamefont {Ozeri}, \citenamefont {Seidelin},\ and\ \citenamefont
  {Wineland}}]{Knill2008}%
  \BibitemOpen
  \bibfield  {author} {\bibinfo {author} {\bibfnamefont {E.}~\bibnamefont
  {Knill}}, \bibinfo {author} {\bibfnamefont {D.}~\bibnamefont {Leibfried}},
  \bibinfo {author} {\bibfnamefont {R.}~\bibnamefont {Reichle}}, \bibinfo
  {author} {\bibfnamefont {J.}~\bibnamefont {Britton}}, \bibinfo {author}
  {\bibfnamefont {R.~B.}\ \bibnamefont {Blakestad}}, \bibinfo {author}
  {\bibfnamefont {J.~D.}\ \bibnamefont {Jost}}, \bibinfo {author}
  {\bibfnamefont {C.}~\bibnamefont {Langer}}, \bibinfo {author} {\bibfnamefont
  {R.}~\bibnamefont {Ozeri}}, \bibinfo {author} {\bibfnamefont
  {S.}~\bibnamefont {Seidelin}},\ and\ \bibinfo {author} {\bibfnamefont
  {D.~J.}\ \bibnamefont {Wineland}},\ }\href
  {https://doi.org/10.1103/PhysRevA.77.012307} {\bibfield  {journal} {\bibinfo
  {journal} {Phys. Rev. A}\ }\textbf {\bibinfo {volume} {77}},\ \bibinfo
  {pages} {012307} (\bibinfo {year} {2008})}\BibitemShut {NoStop}%
\bibitem [{\citenamefont {Dankert}\ \emph {et~al.}(2009)\citenamefont
  {Dankert}, \citenamefont {Cleve}, \citenamefont {Emerson},\ and\
  \citenamefont {Livine}}]{Dankert09}%
  \BibitemOpen
  \bibfield  {author} {\bibinfo {author} {\bibfnamefont {C.}~\bibnamefont
  {Dankert}}, \bibinfo {author} {\bibfnamefont {R.}~\bibnamefont {Cleve}},
  \bibinfo {author} {\bibfnamefont {J.}~\bibnamefont {Emerson}},\ and\ \bibinfo
  {author} {\bibfnamefont {E.}~\bibnamefont {Livine}},\ }\href
  {https://doi.org/10.1103/PhysRevA.80.012304} {\bibfield  {journal} {\bibinfo
  {journal} {Phys. Rev. A}\ }\textbf {\bibinfo {volume} {80}},\ \bibinfo
  {pages} {012304} (\bibinfo {year} {2009})}\BibitemShut {NoStop}%
\bibitem [{\citenamefont {Chow}\ \emph {et~al.}(2009)\citenamefont {Chow},
  \citenamefont {Gambetta}, \citenamefont {Tornberg}, \citenamefont {Koch},
  \citenamefont {Bishop}, \citenamefont {Houck}, \citenamefont {Johnson},
  \citenamefont {Frunzio}, \citenamefont {Girvin},\ and\ \citenamefont
  {Schoelkopf}}]{Chow09}%
  \BibitemOpen
  \bibfield  {author} {\bibinfo {author} {\bibfnamefont {J.~M.}\ \bibnamefont
  {Chow}}, \bibinfo {author} {\bibfnamefont {J.~M.}\ \bibnamefont {Gambetta}},
  \bibinfo {author} {\bibfnamefont {L.}~\bibnamefont {Tornberg}}, \bibinfo
  {author} {\bibfnamefont {J.}~\bibnamefont {Koch}}, \bibinfo {author}
  {\bibfnamefont {L.~S.}\ \bibnamefont {Bishop}}, \bibinfo {author}
  {\bibfnamefont {A.~A.}\ \bibnamefont {Houck}}, \bibinfo {author}
  {\bibfnamefont {B.~R.}\ \bibnamefont {Johnson}}, \bibinfo {author}
  {\bibfnamefont {L.}~\bibnamefont {Frunzio}}, \bibinfo {author} {\bibfnamefont
  {S.~M.}\ \bibnamefont {Girvin}},\ and\ \bibinfo {author} {\bibfnamefont
  {R.~J.}\ \bibnamefont {Schoelkopf}},\ }\href
  {https://doi.org/10.1103/PhysRevLett.102.090502} {\bibfield  {journal}
  {\bibinfo  {journal} {Phys. Rev. Lett.}\ }\textbf {\bibinfo {volume} {102}},\
  \bibinfo {pages} {090502} (\bibinfo {year} {2009})}\BibitemShut {NoStop}%
\bibitem [{\citenamefont {Magesan}\ \emph {et~al.}(2011)\citenamefont
  {Magesan}, \citenamefont {Gambetta},\ and\ \citenamefont
  {Emerson}}]{Magesan11}%
  \BibitemOpen
  \bibfield  {author} {\bibinfo {author} {\bibfnamefont {E.}~\bibnamefont
  {Magesan}}, \bibinfo {author} {\bibfnamefont {J.~M.}\ \bibnamefont
  {Gambetta}},\ and\ \bibinfo {author} {\bibfnamefont {J.}~\bibnamefont
  {Emerson}},\ }\href {https://doi.org/10.1103/PhysRevLett.106.180504}
  {\bibfield  {journal} {\bibinfo  {journal} {Phys. Rev. Lett.}\ }\textbf
  {\bibinfo {volume} {106}},\ \bibinfo {pages} {180504} (\bibinfo {year}
  {2011})}\BibitemShut {NoStop}%
\bibitem [{\citenamefont {Magesan}\ \emph
  {et~al.}(2012{\natexlab{a}})\citenamefont {Magesan}, \citenamefont
  {Gambetta},\ and\ \citenamefont {Emerson}}]{Magesan11_2}%
  \BibitemOpen
  \bibfield  {author} {\bibinfo {author} {\bibfnamefont {E.}~\bibnamefont
  {Magesan}}, \bibinfo {author} {\bibfnamefont {J.~M.}\ \bibnamefont
  {Gambetta}},\ and\ \bibinfo {author} {\bibfnamefont {J.}~\bibnamefont
  {Emerson}},\ }\href {https://doi.org/10.1103/PhysRevA.85.042311} {\bibfield
  {journal} {\bibinfo  {journal} {Phys. Rev. A}\ }\textbf {\bibinfo {volume}
  {85}},\ \bibinfo {pages} {042311} (\bibinfo {year}
  {2012}{\natexlab{a}})}\BibitemShut {NoStop}%
\bibitem [{\citenamefont {Kimmel}\ \emph {et~al.}(2014)\citenamefont {Kimmel},
  \citenamefont {da~Silva}, \citenamefont {Ryan}, \citenamefont {Johnson},\
  and\ \citenamefont {Ohki}}]{Kimmel2014}%
  \BibitemOpen
  \bibfield  {author} {\bibinfo {author} {\bibfnamefont {S.}~\bibnamefont
  {Kimmel}}, \bibinfo {author} {\bibfnamefont {M.~P.}\ \bibnamefont
  {da~Silva}}, \bibinfo {author} {\bibfnamefont {C.~A.}\ \bibnamefont {Ryan}},
  \bibinfo {author} {\bibfnamefont {B.~R.}\ \bibnamefont {Johnson}},\ and\
  \bibinfo {author} {\bibfnamefont {T.}~\bibnamefont {Ohki}},\ }\href
  {https://doi.org/10.1103/PhysRevX.4.011050} {\bibfield  {journal} {\bibinfo
  {journal} {Phys. Rev. X}\ }\textbf {\bibinfo {volume} {4}},\ \bibinfo {pages}
  {011050} (\bibinfo {year} {2014})}\BibitemShut {NoStop}%
\bibitem [{\citenamefont {Epstein}\ \emph {et~al.}(2014)\citenamefont
  {Epstein}, \citenamefont {Cross}, \citenamefont {Magesan},\ and\
  \citenamefont {Gambetta}}]{Epstein14}%
  \BibitemOpen
  \bibfield  {author} {\bibinfo {author} {\bibfnamefont {J.~M.}\ \bibnamefont
  {Epstein}}, \bibinfo {author} {\bibfnamefont {A.~W.}\ \bibnamefont {Cross}},
  \bibinfo {author} {\bibfnamefont {E.}~\bibnamefont {Magesan}},\ and\ \bibinfo
  {author} {\bibfnamefont {J.~M.}\ \bibnamefont {Gambetta}},\ }\href
  {https://doi.org/10.1103/PhysRevA.89.062321} {\bibfield  {journal} {\bibinfo
  {journal} {Phys. Rev. A}\ }\textbf {\bibinfo {volume} {89}},\ \bibinfo
  {pages} {062321} (\bibinfo {year} {2014})}\BibitemShut {NoStop}%
\bibitem [{\citenamefont {Helsen}\ \emph {et~al.}(2020)\citenamefont {Helsen},
  \citenamefont {Roth}, \citenamefont {Onorati}, \citenamefont {Werner},\ and\
  \citenamefont {Eisert}}]{helsen2020general}%
  \BibitemOpen
  \bibfield  {author} {\bibinfo {author} {\bibfnamefont {J.}~\bibnamefont
  {Helsen}}, \bibinfo {author} {\bibfnamefont {I.}~\bibnamefont {Roth}},
  \bibinfo {author} {\bibfnamefont {E.}~\bibnamefont {Onorati}}, \bibinfo
  {author} {\bibfnamefont {A.~H.}\ \bibnamefont {Werner}},\ and\ \bibinfo
  {author} {\bibfnamefont {J.}~\bibnamefont {Eisert}},\ }\href@noop {}
  {\bibfield  {journal} {\bibinfo  {journal} {e-print arXiv: 2010.07974
  [quant-ph]}\ } (\bibinfo {year} {2020})}\BibitemShut {NoStop}%
\bibitem [{\citenamefont {Wallman}\ and\ \citenamefont
  {Flammia}(2014)}]{Wallman2014}%
  \BibitemOpen
  \bibfield  {author} {\bibinfo {author} {\bibfnamefont {J.~J.}\ \bibnamefont
  {Wallman}}\ and\ \bibinfo {author} {\bibfnamefont {S.~T.}\ \bibnamefont
  {Flammia}},\ }\href {https://doi.org/10.1088/1367-2630/16/10/103032}
  {\bibfield  {journal} {\bibinfo  {journal} {New J. Phys.}\ }\textbf {\bibinfo
  {volume} {16}},\ \bibinfo {pages} {103032} (\bibinfo {year}
  {2014})}\BibitemShut {NoStop}%
\bibitem [{\citenamefont {Granade}\ \emph {et~al.}(2015)\citenamefont
  {Granade}, \citenamefont {Ferrie},\ and\ \citenamefont {Cory}}]{Granade15}%
  \BibitemOpen
  \bibfield  {author} {\bibinfo {author} {\bibfnamefont {C.}~\bibnamefont
  {Granade}}, \bibinfo {author} {\bibfnamefont {C.}~\bibnamefont {Ferrie}},\
  and\ \bibinfo {author} {\bibfnamefont {D.~G.}\ \bibnamefont {Cory}},\ }\href
  {https://doi.org/10.1088/1367-2630/17/1/013042} {\bibfield  {journal}
  {\bibinfo  {journal} {New J. Phys.}\ }\textbf {\bibinfo {volume} {17}},\
  \bibinfo {pages} {013042} (\bibinfo {year} {2015})}\BibitemShut {NoStop}%
\bibitem [{\citenamefont {Fran{\c{c}}a}\ and\ \citenamefont
  {Hashagen}(2018)}]{Frana2018}%
  \BibitemOpen
  \bibfield  {author} {\bibinfo {author} {\bibfnamefont {D.~S.}\ \bibnamefont
  {Fran{\c{c}}a}}\ and\ \bibinfo {author} {\bibfnamefont {A.~K.}\ \bibnamefont
  {Hashagen}},\ }\href {https://doi.org/10.1088/1751-8121/aad6fa} {\bibfield
  {journal} {\bibinfo  {journal} {J. Phys. A}\ }\textbf {\bibinfo {volume}
  {51}},\ \bibinfo {pages} {395302} (\bibinfo {year} {2018})}\BibitemShut
  {NoStop}%
\bibitem [{\citenamefont {Roth}\ \emph {et~al.}(2018)\citenamefont {Roth},
  \citenamefont {Kueng}, \citenamefont {Kimmel}, \citenamefont {Liu},
  \citenamefont {Gross}, \citenamefont {Eisert},\ and\ \citenamefont
  {Kliesch}}]{Roth18}%
  \BibitemOpen
  \bibfield  {author} {\bibinfo {author} {\bibfnamefont {I.}~\bibnamefont
  {Roth}}, \bibinfo {author} {\bibfnamefont {R.}~\bibnamefont {Kueng}},
  \bibinfo {author} {\bibfnamefont {S.}~\bibnamefont {Kimmel}}, \bibinfo
  {author} {\bibfnamefont {Y.-K.}\ \bibnamefont {Liu}}, \bibinfo {author}
  {\bibfnamefont {D.}~\bibnamefont {Gross}}, \bibinfo {author} {\bibfnamefont
  {J.}~\bibnamefont {Eisert}},\ and\ \bibinfo {author} {\bibfnamefont
  {M.}~\bibnamefont {Kliesch}},\ }\href
  {https://doi.org/10.1103/PhysRevLett.121.170502} {\bibfield  {journal}
  {\bibinfo  {journal} {Phys. Rev. Lett.}\ }\textbf {\bibinfo {volume} {121}},\
  \bibinfo {pages} {170502} (\bibinfo {year} {2018})}\BibitemShut {NoStop}%
\bibitem [{\citenamefont {Dirkse}\ \emph {et~al.}(2019)\citenamefont {Dirkse},
  \citenamefont {Helsen},\ and\ \citenamefont {Wehner}}]{Dirkse19}%
  \BibitemOpen
  \bibfield  {author} {\bibinfo {author} {\bibfnamefont {B.}~\bibnamefont
  {Dirkse}}, \bibinfo {author} {\bibfnamefont {J.}~\bibnamefont {Helsen}},\
  and\ \bibinfo {author} {\bibfnamefont {S.}~\bibnamefont {Wehner}},\ }\href
  {https://doi.org/10.1103/PhysRevA.99.012315} {\bibfield  {journal} {\bibinfo
  {journal} {Phys. Rev. A}\ }\textbf {\bibinfo {volume} {99}},\ \bibinfo
  {pages} {012315} (\bibinfo {year} {2019})}\BibitemShut {NoStop}%
\bibitem [{\citenamefont {Boone}\ \emph {et~al.}(2019)\citenamefont {Boone},
  \citenamefont {Carignan-Dugas}, \citenamefont {Wallman},\ and\ \citenamefont
  {Emerson}}]{Boone19}%
  \BibitemOpen
  \bibfield  {author} {\bibinfo {author} {\bibfnamefont {K.}~\bibnamefont
  {Boone}}, \bibinfo {author} {\bibfnamefont {A.}~\bibnamefont
  {Carignan-Dugas}}, \bibinfo {author} {\bibfnamefont {J.~J.}\ \bibnamefont
  {Wallman}},\ and\ \bibinfo {author} {\bibfnamefont {J.}~\bibnamefont
  {Emerson}},\ }\href {https://doi.org/10.1103/PhysRevA.99.032329} {\bibfield
  {journal} {\bibinfo  {journal} {Phys. Rev. A}\ }\textbf {\bibinfo {volume}
  {99}},\ \bibinfo {pages} {032329} (\bibinfo {year} {2019})}\BibitemShut
  {NoStop}%
\bibitem [{\citenamefont {Harper}\ \emph {et~al.}(2019)\citenamefont {Harper},
  \citenamefont {Hincks}, \citenamefont {Ferrie}, \citenamefont {Flammia},\
  and\ \citenamefont {Wallman}}]{Harper19}%
  \BibitemOpen
  \bibfield  {author} {\bibinfo {author} {\bibfnamefont {R.}~\bibnamefont
  {Harper}}, \bibinfo {author} {\bibfnamefont {I.}~\bibnamefont {Hincks}},
  \bibinfo {author} {\bibfnamefont {C.}~\bibnamefont {Ferrie}}, \bibinfo
  {author} {\bibfnamefont {S.~T.}\ \bibnamefont {Flammia}},\ and\ \bibinfo
  {author} {\bibfnamefont {J.~J.}\ \bibnamefont {Wallman}},\ }\href
  {https://doi.org/10.1103/PhysRevA.99.052350} {\bibfield  {journal} {\bibinfo
  {journal} {Phys. Rev. A}\ }\textbf {\bibinfo {volume} {99}},\ \bibinfo
  {pages} {052350} (\bibinfo {year} {2019})}\BibitemShut {NoStop}%
\bibitem [{\citenamefont {Helsen}\ \emph
  {et~al.}(2019{\natexlab{a}})\citenamefont {Helsen}, \citenamefont {Wallman},
  \citenamefont {Flammia},\ and\ \citenamefont {Wehner}}]{Helsen19_2}%
  \BibitemOpen
  \bibfield  {author} {\bibinfo {author} {\bibfnamefont {J.}~\bibnamefont
  {Helsen}}, \bibinfo {author} {\bibfnamefont {J.~J.}\ \bibnamefont {Wallman}},
  \bibinfo {author} {\bibfnamefont {S.~T.}\ \bibnamefont {Flammia}},\ and\
  \bibinfo {author} {\bibfnamefont {S.}~\bibnamefont {Wehner}},\ }\href
  {https://doi.org/10.1103/PhysRevA.100.032304} {\bibfield  {journal} {\bibinfo
   {journal} {Phys. Rev. A}\ }\textbf {\bibinfo {volume} {100}},\ \bibinfo
  {pages} {032304} (\bibinfo {year} {2019}{\natexlab{a}})}\BibitemShut
  {NoStop}%
\bibitem [{\citenamefont {Alexander}\ \emph {et~al.}(2016)\citenamefont
  {Alexander}, \citenamefont {Turner},\ and\ \citenamefont
  {Bartlett}}]{Alexander16}%
  \BibitemOpen
  \bibfield  {author} {\bibinfo {author} {\bibfnamefont {R.~N.}\ \bibnamefont
  {Alexander}}, \bibinfo {author} {\bibfnamefont {P.~S.}\ \bibnamefont
  {Turner}},\ and\ \bibinfo {author} {\bibfnamefont {S.~D.}\ \bibnamefont
  {Bartlett}},\ }\href {https://doi.org/10.1103/PhysRevA.94.032303} {\bibfield
  {journal} {\bibinfo  {journal} {Phys. Rev. A}\ }\textbf {\bibinfo {volume}
  {94}},\ \bibinfo {pages} {032303} (\bibinfo {year} {2016})}\BibitemShut
  {NoStop}%
\bibitem [{\citenamefont {Magesan}\ \emph
  {et~al.}(2012{\natexlab{b}})\citenamefont {Magesan}, \citenamefont
  {Gambetta}, \citenamefont {Johnson}, \citenamefont {Ryan}, \citenamefont
  {Chow}, \citenamefont {Merkel}, \citenamefont {da~Silva}, \citenamefont
  {Keefe}, \citenamefont {Rothwell}, \citenamefont {Ohki}, \citenamefont
  {Ketchen},\ and\ \citenamefont {Steffen}}]{Magesan12}%
  \BibitemOpen
  \bibfield  {author} {\bibinfo {author} {\bibfnamefont {E.}~\bibnamefont
  {Magesan}}, \bibinfo {author} {\bibfnamefont {J.~M.}\ \bibnamefont
  {Gambetta}}, \bibinfo {author} {\bibfnamefont {B.~R.}\ \bibnamefont
  {Johnson}}, \bibinfo {author} {\bibfnamefont {C.~A.}\ \bibnamefont {Ryan}},
  \bibinfo {author} {\bibfnamefont {J.~M.}\ \bibnamefont {Chow}}, \bibinfo
  {author} {\bibfnamefont {S.~T.}\ \bibnamefont {Merkel}}, \bibinfo {author}
  {\bibfnamefont {M.~P.}\ \bibnamefont {da~Silva}}, \bibinfo {author}
  {\bibfnamefont {G.~A.}\ \bibnamefont {Keefe}}, \bibinfo {author}
  {\bibfnamefont {M.~B.}\ \bibnamefont {Rothwell}}, \bibinfo {author}
  {\bibfnamefont {T.~A.}\ \bibnamefont {Ohki}}, \bibinfo {author}
  {\bibfnamefont {M.~B.}\ \bibnamefont {Ketchen}},\ and\ \bibinfo {author}
  {\bibfnamefont {M.}~\bibnamefont {Steffen}},\ }\href
  {https://doi.org/10.1103/PhysRevLett.109.080505} {\bibfield  {journal}
  {\bibinfo  {journal} {Phys. Rev. Lett.}\ }\textbf {\bibinfo {volume} {109}},\
  \bibinfo {pages} {080505} (\bibinfo {year} {2012}{\natexlab{b}})}\BibitemShut
  {NoStop}%
\bibitem [{\citenamefont {Harper}\ and\ \citenamefont
  {Flammia}(2017)}]{Harper2017}%
  \BibitemOpen
  \bibfield  {author} {\bibinfo {author} {\bibfnamefont {R.}~\bibnamefont
  {Harper}}\ and\ \bibinfo {author} {\bibfnamefont {S.~T.}\ \bibnamefont
  {Flammia}},\ }\href {https://doi.org/10.1088/2058-9565/aa5f8d} {\bibfield
  {journal} {\bibinfo  {journal} {Quantum Sci. and Technol.}\ }\textbf
  {\bibinfo {volume} {2}},\ \bibinfo {pages} {015008} (\bibinfo {year}
  {2017})}\BibitemShut {NoStop}%
\bibitem [{\citenamefont {Onorati}\ \emph {et~al.}(2019)\citenamefont
  {Onorati}, \citenamefont {Werner},\ and\ \citenamefont {Eisert}}]{Onorati19}%
  \BibitemOpen
  \bibfield  {author} {\bibinfo {author} {\bibfnamefont {E.}~\bibnamefont
  {Onorati}}, \bibinfo {author} {\bibfnamefont {A.~H.}\ \bibnamefont
  {Werner}},\ and\ \bibinfo {author} {\bibfnamefont {J.}~\bibnamefont
  {Eisert}},\ }\href {https://doi.org/10.1103/PhysRevLett.123.060501}
  {\bibfield  {journal} {\bibinfo  {journal} {Phys. Rev. Lett.}\ }\textbf
  {\bibinfo {volume} {123}},\ \bibinfo {pages} {060501} (\bibinfo {year}
  {2019})}\BibitemShut {NoStop}%
\bibitem [{\citenamefont {Helsen}\ \emph
  {et~al.}(2019{\natexlab{b}})\citenamefont {Helsen}, \citenamefont {Xue},
  \citenamefont {Vandersypen},\ and\ \citenamefont {Wehner}}]{Helsen2019}%
  \BibitemOpen
  \bibfield  {author} {\bibinfo {author} {\bibfnamefont {J.}~\bibnamefont
  {Helsen}}, \bibinfo {author} {\bibfnamefont {X.}~\bibnamefont {Xue}},
  \bibinfo {author} {\bibfnamefont {L.~M.~K.}\ \bibnamefont {Vandersypen}},\
  and\ \bibinfo {author} {\bibfnamefont {S.}~\bibnamefont {Wehner}},\ }\href
  {https://doi.org/10.1038/s41534-019-0182-7} {\bibfield  {journal} {\bibinfo
  {journal} {npj Quantum Inf.}\ }\textbf {\bibinfo {volume} {5}},\ \bibinfo
  {pages} {71} (\bibinfo {year} {2019}{\natexlab{b}})}\BibitemShut {NoStop}%
\bibitem [{\citenamefont {Gaebler}\ \emph {et~al.}(2012)\citenamefont
  {Gaebler}, \citenamefont {Meier}, \citenamefont {Tan}, \citenamefont
  {Bowler}, \citenamefont {Lin}, \citenamefont {Hanneke}, \citenamefont {Jost},
  \citenamefont {Home}, \citenamefont {Knill}, \citenamefont {Leibfried},\ and\
  \citenamefont {Wineland}}]{Gaebler12}%
  \BibitemOpen
  \bibfield  {author} {\bibinfo {author} {\bibfnamefont {J.~P.}\ \bibnamefont
  {Gaebler}}, \bibinfo {author} {\bibfnamefont {A.~M.}\ \bibnamefont {Meier}},
  \bibinfo {author} {\bibfnamefont {T.~R.}\ \bibnamefont {Tan}}, \bibinfo
  {author} {\bibfnamefont {R.}~\bibnamefont {Bowler}}, \bibinfo {author}
  {\bibfnamefont {Y.}~\bibnamefont {Lin}}, \bibinfo {author} {\bibfnamefont
  {D.}~\bibnamefont {Hanneke}}, \bibinfo {author} {\bibfnamefont {J.~D.}\
  \bibnamefont {Jost}}, \bibinfo {author} {\bibfnamefont {J.~P.}\ \bibnamefont
  {Home}}, \bibinfo {author} {\bibfnamefont {E.}~\bibnamefont {Knill}},
  \bibinfo {author} {\bibfnamefont {D.}~\bibnamefont {Leibfried}},\ and\
  \bibinfo {author} {\bibfnamefont {D.~J.}\ \bibnamefont {Wineland}},\ }\href
  {https://doi.org/10.1103/PhysRevLett.108.260503} {\bibfield  {journal}
  {\bibinfo  {journal} {Phys. Rev. Lett.}\ }\textbf {\bibinfo {volume} {108}},\
  \bibinfo {pages} {260503} (\bibinfo {year} {2012})}\BibitemShut {NoStop}%
\bibitem [{\citenamefont {C\'orcoles}\ \emph {et~al.}(2013)\citenamefont
  {C\'orcoles}, \citenamefont {Gambetta}, \citenamefont {Chow}, \citenamefont
  {Smolin}, \citenamefont {Ware}, \citenamefont {Strand}, \citenamefont
  {Plourde},\ and\ \citenamefont {Steffen}}]{Corcoles13}%
  \BibitemOpen
  \bibfield  {author} {\bibinfo {author} {\bibfnamefont {A.~D.}\ \bibnamefont
  {C\'orcoles}}, \bibinfo {author} {\bibfnamefont {J.~M.}\ \bibnamefont
  {Gambetta}}, \bibinfo {author} {\bibfnamefont {J.~M.}\ \bibnamefont {Chow}},
  \bibinfo {author} {\bibfnamefont {J.~A.}\ \bibnamefont {Smolin}}, \bibinfo
  {author} {\bibfnamefont {M.}~\bibnamefont {Ware}}, \bibinfo {author}
  {\bibfnamefont {J.}~\bibnamefont {Strand}}, \bibinfo {author} {\bibfnamefont
  {B.~L.~T.}\ \bibnamefont {Plourde}},\ and\ \bibinfo {author} {\bibfnamefont
  {M.}~\bibnamefont {Steffen}},\ }\href
  {https://doi.org/10.1103/PhysRevA.87.030301} {\bibfield  {journal} {\bibinfo
  {journal} {Phys. Rev. A}\ }\textbf {\bibinfo {volume} {87}},\ \bibinfo
  {pages} {030301} (\bibinfo {year} {2013})}\BibitemShut {NoStop}%
\bibitem [{\citenamefont {Xia}\ \emph {et~al.}(2015)\citenamefont {Xia},
  \citenamefont {Lichtman}, \citenamefont {Maller}, \citenamefont {Carr},
  \citenamefont {Piotrowicz}, \citenamefont {Isenhower},\ and\ \citenamefont
  {Saffman}}]{Xia15}%
  \BibitemOpen
  \bibfield  {author} {\bibinfo {author} {\bibfnamefont {T.}~\bibnamefont
  {Xia}}, \bibinfo {author} {\bibfnamefont {M.}~\bibnamefont {Lichtman}},
  \bibinfo {author} {\bibfnamefont {K.}~\bibnamefont {Maller}}, \bibinfo
  {author} {\bibfnamefont {A.~W.}\ \bibnamefont {Carr}}, \bibinfo {author}
  {\bibfnamefont {M.~J.}\ \bibnamefont {Piotrowicz}}, \bibinfo {author}
  {\bibfnamefont {L.}~\bibnamefont {Isenhower}},\ and\ \bibinfo {author}
  {\bibfnamefont {M.}~\bibnamefont {Saffman}},\ }\href
  {https://doi.org/10.1103/PhysRevLett.114.100503} {\bibfield  {journal}
  {\bibinfo  {journal} {Phys. Rev. Lett.}\ }\textbf {\bibinfo {volume} {114}},\
  \bibinfo {pages} {100503} (\bibinfo {year} {2015})}\BibitemShut {NoStop}%
\bibitem [{\citenamefont {Gambetta}\ \emph {et~al.}(2012)\citenamefont
  {Gambetta}, \citenamefont {C\'orcoles}, \citenamefont {Merkel}, \citenamefont
  {Johnson}, \citenamefont {Smolin}, \citenamefont {Chow}, \citenamefont
  {Ryan}, \citenamefont {Rigetti}, \citenamefont {Poletto}, \citenamefont
  {Ohki}, \citenamefont {Ketchen},\ and\ \citenamefont
  {Steffen}}]{Gambetta2012}%
  \BibitemOpen
  \bibfield  {author} {\bibinfo {author} {\bibfnamefont {J.~M.}\ \bibnamefont
  {Gambetta}}, \bibinfo {author} {\bibfnamefont {A.~D.}\ \bibnamefont
  {C\'orcoles}}, \bibinfo {author} {\bibfnamefont {S.~T.}\ \bibnamefont
  {Merkel}}, \bibinfo {author} {\bibfnamefont {B.~R.}\ \bibnamefont {Johnson}},
  \bibinfo {author} {\bibfnamefont {J.~A.}\ \bibnamefont {Smolin}}, \bibinfo
  {author} {\bibfnamefont {J.~M.}\ \bibnamefont {Chow}}, \bibinfo {author}
  {\bibfnamefont {C.~A.}\ \bibnamefont {Ryan}}, \bibinfo {author}
  {\bibfnamefont {C.}~\bibnamefont {Rigetti}}, \bibinfo {author} {\bibfnamefont
  {S.}~\bibnamefont {Poletto}}, \bibinfo {author} {\bibfnamefont {T.~A.}\
  \bibnamefont {Ohki}}, \bibinfo {author} {\bibfnamefont {M.~B.}\ \bibnamefont
  {Ketchen}},\ and\ \bibinfo {author} {\bibfnamefont {M.}~\bibnamefont
  {Steffen}},\ }\href {https://doi.org/10.1103/PhysRevLett.109.240504}
  {\bibfield  {journal} {\bibinfo  {journal} {Phys. Rev. Lett.}\ }\textbf
  {\bibinfo {volume} {109}},\ \bibinfo {pages} {240504} (\bibinfo {year}
  {2012})}\BibitemShut {NoStop}%
\bibitem [{\citenamefont {Carignan-Dugas}\ \emph {et~al.}(2015)\citenamefont
  {Carignan-Dugas}, \citenamefont {Wallman},\ and\ \citenamefont
  {Emerson}}]{Dugas15}%
  \BibitemOpen
  \bibfield  {author} {\bibinfo {author} {\bibfnamefont {A.}~\bibnamefont
  {Carignan-Dugas}}, \bibinfo {author} {\bibfnamefont {J.~J.}\ \bibnamefont
  {Wallman}},\ and\ \bibinfo {author} {\bibfnamefont {J.}~\bibnamefont
  {Emerson}},\ }\href {https://doi.org/10.1103/PhysRevA.92.060302} {\bibfield
  {journal} {\bibinfo  {journal} {Phys. Rev. A}\ }\textbf {\bibinfo {volume}
  {92}},\ \bibinfo {pages} {060302} (\bibinfo {year} {2015})}\BibitemShut
  {NoStop}%
\bibitem [{\citenamefont {Cross}\ \emph {et~al.}(2016)\citenamefont {Cross},
  \citenamefont {Magesan}, \citenamefont {Bishop}, \citenamefont {Smolin},\
  and\ \citenamefont {Gambetta}}]{Cross2016}%
  \BibitemOpen
  \bibfield  {author} {\bibinfo {author} {\bibfnamefont {A.~W.}\ \bibnamefont
  {Cross}}, \bibinfo {author} {\bibfnamefont {E.}~\bibnamefont {Magesan}},
  \bibinfo {author} {\bibfnamefont {L.~S.}\ \bibnamefont {Bishop}}, \bibinfo
  {author} {\bibfnamefont {J.~A.}\ \bibnamefont {Smolin}},\ and\ \bibinfo
  {author} {\bibfnamefont {J.~M.}\ \bibnamefont {Gambetta}},\ }\href
  {https://doi.org/10.1038/npjqi.2016.12} {\bibfield  {journal} {\bibinfo
  {journal} {npj Quantum Inf.}\ }\textbf {\bibinfo {volume} {2}},\ \bibinfo
  {pages} {16012} (\bibinfo {year} {2016})}\BibitemShut {NoStop}%
\bibitem [{\citenamefont {Hashagen}\ \emph {et~al.}(2018)\citenamefont
  {Hashagen}, \citenamefont {Flammia}, \citenamefont {Gross},\ and\
  \citenamefont {Wallman}}]{Hashagen2018}%
  \BibitemOpen
  \bibfield  {author} {\bibinfo {author} {\bibfnamefont {A.~K.}\ \bibnamefont
  {Hashagen}}, \bibinfo {author} {\bibfnamefont {S.~T.}\ \bibnamefont
  {Flammia}}, \bibinfo {author} {\bibfnamefont {D.}~\bibnamefont {Gross}},\
  and\ \bibinfo {author} {\bibfnamefont {J.~J.}\ \bibnamefont {Wallman}},\
  }\href {https://doi.org/10.22331/q-2018-08-22-85} {\bibfield  {journal}
  {\bibinfo  {journal} {{Quantum}}\ }\textbf {\bibinfo {volume} {2}},\ \bibinfo
  {pages} {85} (\bibinfo {year} {2018})}\BibitemShut {NoStop}%
\bibitem [{\citenamefont {Brown}\ and\ \citenamefont {Eastin}(2018)}]{Brown18}%
  \BibitemOpen
  \bibfield  {author} {\bibinfo {author} {\bibfnamefont {W.~G.}\ \bibnamefont
  {Brown}}\ and\ \bibinfo {author} {\bibfnamefont {B.}~\bibnamefont {Eastin}},\
  }\href {https://doi.org/10.1103/PhysRevA.97.062323} {\bibfield  {journal}
  {\bibinfo  {journal} {Phys. Rev. A}\ }\textbf {\bibinfo {volume} {97}},\
  \bibinfo {pages} {062323} (\bibinfo {year} {2018})}\BibitemShut {NoStop}%
\bibitem [{\citenamefont {Erhard}\ \emph {et~al.}(2019)\citenamefont {Erhard},
  \citenamefont {Wallman}, \citenamefont {Postler}, \citenamefont {Meth},
  \citenamefont {Stricker}, \citenamefont {Martinez}, \citenamefont
  {Schindler}, \citenamefont {Monz}, \citenamefont {Emerson},\ and\
  \citenamefont {Blatt}}]{Erhard2019}%
  \BibitemOpen
  \bibfield  {author} {\bibinfo {author} {\bibfnamefont {A.}~\bibnamefont
  {Erhard}}, \bibinfo {author} {\bibfnamefont {J.~J.}\ \bibnamefont {Wallman}},
  \bibinfo {author} {\bibfnamefont {L.}~\bibnamefont {Postler}}, \bibinfo
  {author} {\bibfnamefont {M.}~\bibnamefont {Meth}}, \bibinfo {author}
  {\bibfnamefont {R.}~\bibnamefont {Stricker}}, \bibinfo {author}
  {\bibfnamefont {E.~A.}\ \bibnamefont {Martinez}}, \bibinfo {author}
  {\bibfnamefont {P.}~\bibnamefont {Schindler}}, \bibinfo {author}
  {\bibfnamefont {T.}~\bibnamefont {Monz}}, \bibinfo {author} {\bibfnamefont
  {J.}~\bibnamefont {Emerson}},\ and\ \bibinfo {author} {\bibfnamefont
  {R.}~\bibnamefont {Blatt}},\ }\href
  {https://doi.org/10.1038/s41467-019-13068-7} {\bibfield  {journal} {\bibinfo
  {journal} {Nat. Commun.}\ }\textbf {\bibinfo {volume} {10}},\ \bibinfo
  {pages} {5347} (\bibinfo {year} {2019})}\BibitemShut {NoStop}%
\bibitem [{\citenamefont {Zhou}\ \emph {et~al.}(2011)\citenamefont {Zhou},
  \citenamefont {Ralph}, \citenamefont {Kalasuwan}, \citenamefont {Zhang},
  \citenamefont {Peruzzo}, \citenamefont {Lanyon},\ and\ \citenamefont
  {O'brien}}]{Zhou2011}%
  \BibitemOpen
  \bibfield  {author} {\bibinfo {author} {\bibfnamefont {X.-Q.}\ \bibnamefont
  {Zhou}}, \bibinfo {author} {\bibfnamefont {T.~C.}\ \bibnamefont {Ralph}},
  \bibinfo {author} {\bibfnamefont {P.}~\bibnamefont {Kalasuwan}}, \bibinfo
  {author} {\bibfnamefont {M.}~\bibnamefont {Zhang}}, \bibinfo {author}
  {\bibfnamefont {A.}~\bibnamefont {Peruzzo}}, \bibinfo {author} {\bibfnamefont
  {B.~P.}\ \bibnamefont {Lanyon}},\ and\ \bibinfo {author} {\bibfnamefont
  {J.~L.}\ \bibnamefont {O'brien}},\ }\href
  {https://doi.org/10.1038/ncomms1392} {\bibfield  {journal} {\bibinfo
  {journal} {Nat. Commun.}\ }\textbf {\bibinfo {volume} {2}},\ \bibinfo {pages}
  {413} (\bibinfo {year} {2011})}\BibitemShut {NoStop}%
\bibitem [{\citenamefont {Ara{\'{u}}jo}\ \emph {et~al.}(2014)\citenamefont
  {Ara{\'{u}}jo}, \citenamefont {Feix}, \citenamefont {Costa},\ and\
  \citenamefont {Brukner}}]{Arajo2014}%
  \BibitemOpen
  \bibfield  {author} {\bibinfo {author} {\bibfnamefont {M.}~\bibnamefont
  {Ara{\'{u}}jo}}, \bibinfo {author} {\bibfnamefont {A.}~\bibnamefont {Feix}},
  \bibinfo {author} {\bibfnamefont {F.}~\bibnamefont {Costa}},\ and\ \bibinfo
  {author} {\bibfnamefont {{\v{C}}.}~\bibnamefont {Brukner}},\ }\href
  {https://doi.org/10.1088/1367-2630/16/9/093026} {\bibfield  {journal}
  {\bibinfo  {journal} {New J. Phys.}\ }\textbf {\bibinfo {volume} {16}},\
  \bibinfo {pages} {093026} (\bibinfo {year} {2014})}\BibitemShut {NoStop}%
\bibitem [{\citenamefont {Friis}\ \emph {et~al.}(2014)\citenamefont {Friis},
  \citenamefont {Dunjko}, \citenamefont {D\"ur},\ and\ \citenamefont
  {Briegel}}]{Friis2014}%
  \BibitemOpen
  \bibfield  {author} {\bibinfo {author} {\bibfnamefont {N.}~\bibnamefont
  {Friis}}, \bibinfo {author} {\bibfnamefont {V.}~\bibnamefont {Dunjko}},
  \bibinfo {author} {\bibfnamefont {W.}~\bibnamefont {D\"ur}},\ and\ \bibinfo
  {author} {\bibfnamefont {H.~J.}\ \bibnamefont {Briegel}},\ }\href
  {https://doi.org/10.1103/PhysRevA.89.030303} {\bibfield  {journal} {\bibinfo
  {journal} {Phys. Rev. A}\ }\textbf {\bibinfo {volume} {89}},\ \bibinfo
  {pages} {030303} (\bibinfo {year} {2014})}\BibitemShut {NoStop}%
\bibitem [{\citenamefont {Schmidt-Kaler}\ \emph {et~al.}(2003)\citenamefont
  {Schmidt-Kaler}, \citenamefont {H\"{a}ffner}, \citenamefont {Riebe},
  \citenamefont {Gulde}, \citenamefont {Lancaster}, \citenamefont {Deuschle},
  \citenamefont {Becher}, \citenamefont {Roos}, \citenamefont {Eschner},\ and\
  \citenamefont {Blatt}}]{SchmidtKaler2003}%
  \BibitemOpen
  \bibfield  {author} {\bibinfo {author} {\bibfnamefont {F.}~\bibnamefont
  {Schmidt-Kaler}}, \bibinfo {author} {\bibfnamefont {H.}~\bibnamefont
  {H\"{a}ffner}}, \bibinfo {author} {\bibfnamefont {M.}~\bibnamefont {Riebe}},
  \bibinfo {author} {\bibfnamefont {S.}~\bibnamefont {Gulde}}, \bibinfo
  {author} {\bibfnamefont {G.~P.~T.}\ \bibnamefont {Lancaster}}, \bibinfo
  {author} {\bibfnamefont {T.}~\bibnamefont {Deuschle}}, \bibinfo {author}
  {\bibfnamefont {C.}~\bibnamefont {Becher}}, \bibinfo {author} {\bibfnamefont
  {C.~F.}\ \bibnamefont {Roos}}, \bibinfo {author} {\bibfnamefont
  {J.}~\bibnamefont {Eschner}},\ and\ \bibinfo {author} {\bibfnamefont
  {R.}~\bibnamefont {Blatt}},\ }\href {https://doi.org/10.1038/nature01494}
  {\bibfield  {journal} {\bibinfo  {journal} {Nature}\ }\textbf {\bibinfo
  {volume} {422}},\ \bibinfo {pages} {408} (\bibinfo {year}
  {2003})}\BibitemShut {NoStop}%
\bibitem [{\citenamefont {Barreiro}\ \emph {et~al.}(2011)\citenamefont
  {Barreiro}, \citenamefont {M\"{u}ller}, \citenamefont {Schindler},
  \citenamefont {Nigg}, \citenamefont {Monz}, \citenamefont {Chwalla},
  \citenamefont {Hennrich}, \citenamefont {Roos}, \citenamefont {Zoller},\ and\
  \citenamefont {Blatt}}]{Barreiro2011}%
  \BibitemOpen
  \bibfield  {author} {\bibinfo {author} {\bibfnamefont {J.~T.}\ \bibnamefont
  {Barreiro}}, \bibinfo {author} {\bibfnamefont {M.}~\bibnamefont
  {M\"{u}ller}}, \bibinfo {author} {\bibfnamefont {P.}~\bibnamefont
  {Schindler}}, \bibinfo {author} {\bibfnamefont {D.}~\bibnamefont {Nigg}},
  \bibinfo {author} {\bibfnamefont {T.}~\bibnamefont {Monz}}, \bibinfo {author}
  {\bibfnamefont {M.}~\bibnamefont {Chwalla}}, \bibinfo {author} {\bibfnamefont
  {M.}~\bibnamefont {Hennrich}}, \bibinfo {author} {\bibfnamefont {C.~F.}\
  \bibnamefont {Roos}}, \bibinfo {author} {\bibfnamefont {P.}~\bibnamefont
  {Zoller}},\ and\ \bibinfo {author} {\bibfnamefont {R.}~\bibnamefont
  {Blatt}},\ }\href {https://doi.org/10.1038/nature09801} {\bibfield  {journal}
  {\bibinfo  {journal} {Nature}\ }\textbf {\bibinfo {volume} {470}},\ \bibinfo
  {pages} {486} (\bibinfo {year} {2011})}\BibitemShut {NoStop}%
\bibitem [{\citenamefont {Miguel-Ramiro}\ \emph {et~al.}(2020)\citenamefont
  {Miguel-Ramiro}, \citenamefont {Pirker},\ and\ \citenamefont
  {D{\"u}r}}]{GenuineQN}%
  \BibitemOpen
  \bibfield  {author} {\bibinfo {author} {\bibfnamefont {J.}~\bibnamefont
  {Miguel-Ramiro}}, \bibinfo {author} {\bibfnamefont {A.}~\bibnamefont
  {Pirker}},\ and\ \bibinfo {author} {\bibfnamefont {W.}~\bibnamefont
  {D{\"u}r}},\ }\href@noop {} {\bibfield  {journal} {\bibinfo  {journal}
  {e-print arXiv: 2005.00020 [quant-ph]}\ } (\bibinfo {year}
  {2020})}\BibitemShut {NoStop}%
\bibitem [{\citenamefont {Chiribella}\ \emph {et~al.}(2013)\citenamefont
  {Chiribella}, \citenamefont {D'Ariano}, \citenamefont {Perinotti},\ and\
  \citenamefont {Valiron}}]{Chiribella2013}%
  \BibitemOpen
  \bibfield  {author} {\bibinfo {author} {\bibfnamefont {G.}~\bibnamefont
  {Chiribella}}, \bibinfo {author} {\bibfnamefont {G.~M.}\ \bibnamefont
  {D'Ariano}}, \bibinfo {author} {\bibfnamefont {P.}~\bibnamefont
  {Perinotti}},\ and\ \bibinfo {author} {\bibfnamefont {B.}~\bibnamefont
  {Valiron}},\ }\href {https://doi.org/10.1103/PhysRevA.88.022318} {\bibfield
  {journal} {\bibinfo  {journal} {Phys. Rev. A}\ }\textbf {\bibinfo {volume}
  {88}},\ \bibinfo {pages} {022318} (\bibinfo {year} {2013})}\BibitemShut
  {NoStop}%
\bibitem [{\citenamefont {Ara\'ujo}\ \emph {et~al.}(2014)\citenamefont
  {Ara\'ujo}, \citenamefont {Costa},\ and\ \citenamefont
  {Brukner}}]{Araujo2014}%
  \BibitemOpen
  \bibfield  {author} {\bibinfo {author} {\bibfnamefont {M.}~\bibnamefont
  {Ara\'ujo}}, \bibinfo {author} {\bibfnamefont {F.}~\bibnamefont {Costa}},\
  and\ \bibinfo {author} {\bibfnamefont {{\v{C}}.}~\bibnamefont {Brukner}},\
  }\href {https://doi.org/10.1103/PhysRevLett.113.250402} {\bibfield  {journal}
  {\bibinfo  {journal} {Phys. Rev. Lett.}\ }\textbf {\bibinfo {volume} {113}},\
  \bibinfo {pages} {250402} (\bibinfo {year} {2014})}\BibitemShut {NoStop}%
\bibitem [{\citenamefont {Procopio}\ \emph {et~al.}(2015)\citenamefont
  {Procopio}, \citenamefont {Moqanaki}, \citenamefont {Ara{\'{u}}jo},
  \citenamefont {Costa}, \citenamefont {Calafell}, \citenamefont {Dowd},
  \citenamefont {Hamel}, \citenamefont {Rozema}, \citenamefont {Brukner},\ and\
  \citenamefont {Walther}}]{Procopio2015}%
  \BibitemOpen
  \bibfield  {author} {\bibinfo {author} {\bibfnamefont {L.~M.}\ \bibnamefont
  {Procopio}}, \bibinfo {author} {\bibfnamefont {A.}~\bibnamefont {Moqanaki}},
  \bibinfo {author} {\bibfnamefont {M.}~\bibnamefont {Ara{\'{u}}jo}}, \bibinfo
  {author} {\bibfnamefont {F.}~\bibnamefont {Costa}}, \bibinfo {author}
  {\bibfnamefont {I.~A.}\ \bibnamefont {Calafell}}, \bibinfo {author}
  {\bibfnamefont {E.~G.}\ \bibnamefont {Dowd}}, \bibinfo {author}
  {\bibfnamefont {D.~R.}\ \bibnamefont {Hamel}}, \bibinfo {author}
  {\bibfnamefont {L.~A.}\ \bibnamefont {Rozema}}, \bibinfo {author}
  {\bibfnamefont {{\v{C}}.}~\bibnamefont {Brukner}},\ and\ \bibinfo {author}
  {\bibfnamefont {P.}~\bibnamefont {Walther}},\ }\href
  {https://doi.org/10.1038/ncomms8913} {\bibfield  {journal} {\bibinfo
  {journal} {Nat. Commun.}\ }\textbf {\bibinfo {volume} {6}},\ \bibinfo {pages}
  {7913} (\bibinfo {year} {2015})}\BibitemShut {NoStop}%
\bibitem [{\citenamefont {Carignan-Dugas}\ \emph {et~al.}(2019)\citenamefont
  {Carignan-Dugas}, \citenamefont {Wallman},\ and\ \citenamefont
  {Emerson}}]{Carignan_Dugas_2019}%
  \BibitemOpen
  \bibfield  {author} {\bibinfo {author} {\bibfnamefont {A.}~\bibnamefont
  {Carignan-Dugas}}, \bibinfo {author} {\bibfnamefont {J.~J.}\ \bibnamefont
  {Wallman}},\ and\ \bibinfo {author} {\bibfnamefont {J.}~\bibnamefont
  {Emerson}},\ }\href {https://doi.org/10.1088/1367-2630/ab1800} {\bibfield
  {journal} {\bibinfo  {journal} {New J. Phys.}\ }\textbf {\bibinfo {volume}
  {21}},\ \bibinfo {pages} {053016} (\bibinfo {year} {2019})}\BibitemShut
  {NoStop}%
\bibitem [{\citenamefont {Lin}\ and\ \citenamefont {He}(2015)}]{Lin2015}%
  \BibitemOpen
  \bibfield  {author} {\bibinfo {author} {\bibfnamefont {Q.}~\bibnamefont
  {Lin}}\ and\ \bibinfo {author} {\bibfnamefont {B.}~\bibnamefont {He}},\
  }\href@noop {} {\bibfield  {journal} {\bibinfo  {journal} {Sci. Rep.}\
  }\textbf {\bibinfo {volume} {5}},\ \bibinfo {pages} {12792} (\bibinfo {year}
  {2015})}\BibitemShut {NoStop}%
\bibitem [{\citenamefont {Friis}\ \emph {et~al.}(2015)\citenamefont {Friis},
  \citenamefont {Melnikov}, \citenamefont {Kirchmair},\ and\ \citenamefont
  {Briegel}}]{Friis2015S}%
  \BibitemOpen
  \bibfield  {author} {\bibinfo {author} {\bibfnamefont {N.}~\bibnamefont
  {Friis}}, \bibinfo {author} {\bibfnamefont {A.~A.}\ \bibnamefont {Melnikov}},
  \bibinfo {author} {\bibfnamefont {G.}~\bibnamefont {Kirchmair}},\ and\
  \bibinfo {author} {\bibfnamefont {H.~J.}\ \bibnamefont {Briegel}},\ }\href
  {https://doi.org/10.1038/srep18036} {\bibfield  {journal} {\bibinfo
  {journal} {Sci. Rep.}\ }\textbf {\bibinfo {volume} {5}},\ \bibinfo {pages}
  {18036} (\bibinfo {year} {2015})}\BibitemShut {NoStop}%
\bibitem [{\citenamefont {Rubino}\ \emph {et~al.}(2020)\citenamefont {Rubino},
  \citenamefont {Rozema}, \citenamefont {Ebler}, \citenamefont {Kristjánsson},
  \citenamefont {Salek}, \citenamefont {Guérin}, \citenamefont {Abbott},
  \citenamefont {Branciard}, \citenamefont {Brukner}, \citenamefont
  {Chiribella},\ and\ \citenamefont {Walther}}]{rubino2020}%
  \BibitemOpen
  \bibfield  {author} {\bibinfo {author} {\bibfnamefont {G.}~\bibnamefont
  {Rubino}}, \bibinfo {author} {\bibfnamefont {L.~A.}\ \bibnamefont {Rozema}},
  \bibinfo {author} {\bibfnamefont {D.}~\bibnamefont {Ebler}}, \bibinfo
  {author} {\bibfnamefont {H.}~\bibnamefont {Kristjánsson}}, \bibinfo {author}
  {\bibfnamefont {S.}~\bibnamefont {Salek}}, \bibinfo {author} {\bibfnamefont
  {P.~A.}\ \bibnamefont {Guérin}}, \bibinfo {author} {\bibfnamefont {A.~A.}\
  \bibnamefont {Abbott}}, \bibinfo {author} {\bibfnamefont {C.}~\bibnamefont
  {Branciard}}, \bibinfo {author} {\bibfnamefont {{\v{C}}.}~\bibnamefont
  {Brukner}}, \bibinfo {author} {\bibfnamefont {G.}~\bibnamefont
  {Chiribella}},\ and\ \bibinfo {author} {\bibfnamefont {P.}~\bibnamefont
  {Walther}},\ }\href@noop {} {\bibfield  {journal} {\bibinfo  {journal}
  {e-print arXiv: 2007.05005 [quant-ph]}\ } (\bibinfo {year}
  {2020})}\BibitemShut {NoStop}%
\bibitem [{\citenamefont {Cirac}\ and\ \citenamefont
  {Zoller}(1995)}]{CiracZoller}%
  \BibitemOpen
  \bibfield  {author} {\bibinfo {author} {\bibfnamefont {J.~I.}\ \bibnamefont
  {Cirac}}\ and\ \bibinfo {author} {\bibfnamefont {P.}~\bibnamefont {Zoller}},\
  }\href {https://doi.org/10.1103/PhysRevLett.74.4091} {\bibfield  {journal}
  {\bibinfo  {journal} {Phys. Rev. Lett.}\ }\textbf {\bibinfo {volume} {74}},\
  \bibinfo {pages} {4091} (\bibinfo {year} {1995})}\BibitemShut {NoStop}%
\bibitem [{\citenamefont {Dong}\ \emph {et~al.}(2019)\citenamefont {Dong},
  \citenamefont {Nakayama}, \citenamefont {Soeda},\ and\ \citenamefont
  {Murao}}]{combs20}%
  \BibitemOpen
  \bibfield  {author} {\bibinfo {author} {\bibfnamefont {Q.}~\bibnamefont
  {Dong}}, \bibinfo {author} {\bibfnamefont {S.}~\bibnamefont {Nakayama}},
  \bibinfo {author} {\bibfnamefont {A.}~\bibnamefont {Soeda}},\ and\ \bibinfo
  {author} {\bibfnamefont {M.}~\bibnamefont {Murao}},\ }\href@noop {}
  {\bibfield  {journal} {\bibinfo  {journal} {e-print arXiv: 1911.01645
  [quant-ph]}\ } (\bibinfo {year} {2019})}\BibitemShut {NoStop}%
\bibitem [{\citenamefont {Dankert}(2005)}]{dankert2005e}%
  \BibitemOpen
  \bibfield  {author} {\bibinfo {author} {\bibfnamefont {C.}~\bibnamefont
  {Dankert}},\ }\href@noop {} {\bibfield  {journal} {\bibinfo  {journal}
  {e-print arXiv: 0512217 [quant-ph]}\ } (\bibinfo {year} {2005})}\BibitemShut
  {NoStop}%
\bibitem [{\citenamefont {Lidl}\ and\ \citenamefont
  {Niederreiter}(1994)}]{Lidl1994}%
  \BibitemOpen
  \bibfield  {author} {\bibinfo {author} {\bibfnamefont {R.}~\bibnamefont
  {Lidl}}\ and\ \bibinfo {author} {\bibfnamefont {H.}~\bibnamefont
  {Niederreiter}},\ }\href {https://doi.org/10.1017/cbo9781139172769} {\emph
  {\bibinfo {title} {Introduction to Finite Fields and their Applications}}}\
  (\bibinfo  {publisher} {Cambridge University Press},\ \bibinfo {year}
  {1994})\BibitemShut {NoStop}%
\bibitem [{\citenamefont {M\o{}lmer}\ and\ \citenamefont
  {S\o{}rensen}(1999)}]{MolmerSor}%
  \BibitemOpen
  \bibfield  {author} {\bibinfo {author} {\bibfnamefont {K.}~\bibnamefont
  {M\o{}lmer}}\ and\ \bibinfo {author} {\bibfnamefont {A.}~\bibnamefont
  {S\o{}rensen}},\ }\href {https://doi.org/10.1103/PhysRevLett.82.1835}
  {\bibfield  {journal} {\bibinfo  {journal} {Phys. Rev. Lett.}\ }\textbf
  {\bibinfo {volume} {82}},\ \bibinfo {pages} {1835} (\bibinfo {year}
  {1999})}\BibitemShut {NoStop}%
\end{thebibliography}%

\onecolumngrid
\appendix*

\numberwithin{equation}{section}
\setcounter{equation}{0}
\renewcommand\theequation{A.\arabic{equation}}

\newpage

\subsection*{Appendix A: Noise model and channel matrix}

Each application of a controlled operation introduces
some noise. We consider uncorrelated noise for the control and main
registers, justified by the proposed physical realizations. Assuming no noise in the control system for the moment, the error
can hence be described in a controlled way at each sequence position
$j$,

\begin{equation}
{\xi}^{(j)}(\rho)=\sum_{s}M_{s}^{(j)}\rho M_{s}^{(j)^{\dagger}},\label{eq:map1-1}
\end{equation}
where $M_{s}$ are the global Kraus operators defined in accordance
to \cite{combs20}, i.e.
\begin{equation}
M_{s}^{(j)}=\sum_{i=0}^{k-1}\left|i\right\rangle _{c}\left\langle i\right|\otimes K_{i,s}^{(j)},\label{eq:globalkraus-1}
\end{equation}
where $K_{i,s}^{(j)}$ are the Kraus operators of each gate $i$ of
each branch, and at each sequence position $j$. It is straightforward
to see that, given $\sum_{s}K_{i,s}^{(j)^{\dagger}}K_{i,s}^{(j)}=I$
for every $i$, it implies that $\sum_{s}M_{s}^{(j)^{\dagger}}M_{s}^{(j)}=I$.
For simplicity, in this work we restrict ourselves to the case where
the noise is independent of the sequence branch and position, i.e.
$K_{i,s}^{(j)}=K_{s}$, also known as zeroth-order approximation (see e.g.\cite{Magesan11_2}).
The uncorrelated error on the control state is well justified
from the proposed experimental implementations, where control
is added by external devices. Kraus operators
are hence reduced to
\begin{equation}
M_{s}=I_{d}\otimes K_{s}.\label{eq:finalkraus-1}
\end{equation}
We can equivalently express the noise affecting the main register
by using the Pauli basis decomposition of the Kraus operators, as a
function of the channel matrix $\chi$:
\begin{equation}
{\xi}(\rho)=I_{d}\otimes\sum_{i,j}\chi_{ij}\mathcal{\mathrm{P}}_{i}\rho\mathcal{\mathrm{P}}_{j}^{\dagger},\label{eq:cptp-1}
\end{equation}
where $\chi_{ij}$ are the elements of the channel matrix $\chi$
and $\mathcal{\mathrm{P}}_{i}$ are the $n-$qudit Pauli elements $\mathcal{\mathrm{P}}_{i} \in \left\{X^{i_{1}}Z^{j_{1}}\otimes\cdots\otimes X^{i_{n}}Z^{j_{n}}\right\} $
(see next Appendix B).

\setcounter{equation}{0}
\renewcommand\theequation{B.\arabic{equation}}

\subsection*{Appendix B: Different gate-set instances that coherent RB can benchmark }

In this section, we provide simple proofs of particular instances of sets of quatum gates $U_{i} \in G$  that fulfill
\begin{equation}
\sum_{i=1}^{|G|}U_{i}^{\dagger}P_{\mathbf{j}}U_{i}=\begin{cases}
|G|I_{d}^{\otimes n} & \mathbf{j}=\mathbf{o}\\
0 & \mathbf{j}\neq\mathbf{o}
\end{cases},\label{eq: conditionAp}
\end{equation}
for any Pauli element
$P_{\mathbf{j}}\equiv X^{\mathbf{j}_{r}}Z^{\mathbf{j}_{s}}=X^{s_{1}}Z^{r_{1}}\otimes\cdots\otimes X^{s_{n}}Z^{r_{n}}$, and
therefore can be benchmarked using our coherent approach. Observe that condition Eq. (\ref{eq: conditionAp}) is a much less demanding restriction for a set of gates than the $2-$design condition \cite{Dankert09}.

\subsubsection*{$n-$qudit Pauli operators}

Consider the set of $n-$qudit Pauli operators
\begin{equation}
\mathcal{P}_{d,n}=\left\{ X^{i_{1}}Z^{j_{1}}\otimes\cdots\otimes X^{i_{n}}Z^{j_{n}}\right\}, \label{eq:pauliset}
\end{equation}
with $i_{s},j_{s}\in\mathbb{Z}_{d}$ and where $X,Z$ are the generalized
Pauli operators, i.e.
\begin{equation}
X^{r}:\left|s\right\rangle \longmapsto\left|s\oplus r\right\rangle \,,\,Z^{r}:\left|s\right\rangle \longmapsto w^{rs}\left|s\right\rangle ,\label{eq:pauliops}
\end{equation}
where $w=e^{\frac{2\pi i}{d}}$ and $\oplus$ denotes addition modulo
$d$. We can simplify the notation by defining each element of the
Pauli set as

\begin{equation}
P_{\mathbf{x}}\equiv X^{\mathbf{x}_{i}}Z^{\mathbf{x}_{j}}=X^{i_{1}}Z^{j_{1}}\otimes\cdots\otimes X^{i_{n}}Z^{j_{n}},\label{eq:paulielement}
\end{equation}
where $\mathbf{x}=\left(\mathbf{x}_{i},\mathbf{x}_{j}\right)\in\mathbb{Z}_{d}^{2n}$.
The commutation relation between two elements of the form of Eq. (\ref{eq:paulielement})
is given by
\begin{equation}
P_{\mathbf{x}}P_{\mathbf{y}}=w^{(\mathbf{x},\mathbf{y})_{S_{P}}}P_{\mathbf{y}}P_{\mathbf{x}},\label{eq:comm}
\end{equation}
where $(\mathbf{x},\mathbf{y})_{S_{P}}$ is the symplectic inner product
defined as $(\mathbf{x},\mathbf{y})_{S_{P}}=\mathbf{x}_{i}\cdot\mathbf{y}_{j}-\mathbf{x}_{j}\cdot\mathbf{y}_{i}$. Finally, define $\chi_{\mathbf{q}}$ as the character of $\mathcal{P}$
for any $\mathbf{q}$, such that $\chi_{\mathbf{q}}\left(P_{\mathbf{x}}\right)=\text{\ensuremath{w^{(\mathbf{q},\mathbf{x})_{S_{P}}}}}$.
It follows that, for all $\mathbf{q}\in\mathbb{Z}_{d}^{2n}$, $\mathbf{q}\neq\mathbf{o}$ \cite{dankert2005e,Lidl1994}:
\begin{equation}
\sum_{\mathbf{x}}\chi_{\mathbf{q}}\left(P_{\mathbf{x}}\right)=\text{\ensuremath{w^{(\mathbf{q},\mathbf{x})_{S_{P}}}}}=0.\label{eq:character}
\end{equation}
Consider now Eq. (\ref{eq: conditionAp}),
fulfilled in case the coherent RB
protocol can be applied,
i.e.
\begin{equation}
\frac{1}{|G|}\sum_{\mathbf{i}}P_{\mathbf{i}}^{\dagger}P_{\mathbf{j}}P_{\mathbf{i}}=\begin{cases}
I_{d}^{\otimes n} & \mathbf{j}=\mathbf{o}\\
0 & \mathbf{j}\neq\mathbf{o},
\end{cases}\label{eq:conditionapendix}
\end{equation}
where $|G|=d^{2n}$ is the size (number of elements $\mathbf{i}$)
of the Pauli set.
For $\mathbf{j}=\mathbf{o}$, we have $P_{\mathbf{j}}=I\otimes\cdots\otimes I$
and the condition is trivially fulfilled. In case $\mathbf{j}\neq\mathbf{o}$,
we can apply the commutation relation of Eq. (\ref{eq:comm}), such
that
\begin{equation}
\frac{1}{|G|}\sum_{\mathbf{i}}P_{\mathbf{i}}^{\dagger}P_{\mathbf{j}}P_{\mathbf{i}}=\frac{1}{|G|}\sum_{\mathbf{i}}\text{\ensuremath{w^{(\mathbf{i},\mathbf{j})_{S_{P}}}}}P_{\mathbf{j}}=0\,\,\forall\,\mathbf{j}\neq\mathbf{o},
\label{eq:paulisproofs}
\end{equation}
which follows from property Eq. (\ref{eq:character}). Therefore,
we conclude that the set of $n-$qudit Pauli operators fulfills
Eq.(\ref{eq: conditionAp}) and can be benchmarked with coherent RB.

\subsubsection*{$n-$qudit Clifford group}

The $n-$qudit Clifford group is defined as the normalizer of the
$n-$qudit Pauli operators under conjugation, i.e.
\begin{equation}
\mathcal{C}_{d,n}=\left\{ C\in U\left(d^{n}\right)\mid C\mathcal{P}_{d,n}C^{\dagger}\subseteq\mathcal{P}_{d,n}\right\} /I_{d}^{\otimes n}
\end{equation}
It follows from the definition that, for a fixed
Pauli element $P_{\mathbf{x}}$ (Eq. \ref{eq:paulielement}), the
application of different Clifford operations under conjugation leads  to  equally distributed
operators of the form $w^{r}\mathcal{P}$, with $r\in\left\{ 1,\ldots,d\right\}$.
Given the sum of Eq. (\ref{eq: conditionAp}) and the fact that $\sum_{r=1}^{d}w^{r}=\sum_{r=1}^{d}e^{\frac{2\pi ir}{d}}=0,$
it is straightforward to see that the Clifford group also fulfills Eq. (\ref{eq: conditionAp}).

\subsubsection*{Toffoli gate and multi-controlled qudit operations}

Different sets of controlled operations that can be benchmarked with coherent RB exist, as we now demonstrate. In particular,
we show an example for one control qubit and one target qudit, but the scheme can be extended to an arbitrary number of
control and target qudits. Consider the set of quantum controlled-operations $T_{\mathbf{i,r,s}}\in\mathcal{T}$,
where $\mathbf{\left\{ i,r,s\right\} }\in\mathbb{Z}_{d}^{2}$, such that
\begin{equation}
T_{\mathbf{i,r,s}}=\left(P_{\mathbf{i}}\otimes I\right)\left(\left|0\right\rangle _{c}\left\langle 0\right|\otimes P_{\mathbf{r}}+\left|1\right\rangle _{c}\left\langle 1\right|\otimes P_{\mathbf{s}}\right),
\end{equation}
where $P_{\mathbf{j}}$ is given by Eq. (\ref{eq:paulielement}) with
$n=1$. Consider Eq. (\ref{eq: conditionAp}) for some fixed
$\mathbf{\left\{ k,k'\right\} }\in\mathbb{Z}_{d}^{2}$, i.e.
\begin{equation}
\frac{1}{|G|}\sum_{\mathbf{i,r,s}}T_{\mathbf{i,r,s}}^{\dagger}\left(P_{\mathbf{k}}\otimes P_{\mathbf{k'}}\right)T_{\mathbf{i,r,s}}.
\end{equation}
For $\mathbf{k},\mathbf{k'}=0$, Eq. (\ref{eq: conditionAp}) is trivially fulfilled. For the case $\mathbf{k},\mathbf{k'}\neq 0$, one can expand this expression such that
\small
\begin{equation}
\frac{1}{|G|}\sum_{\mathbf{i,r,s}}\left[\left(P_{\mathbf{i}}^{\dagger}P_{\mathbf{k}}^{00}\left|0\right\rangle \left\langle 0\right|P_{\mathbf{i}}\otimes P_{\mathbf{r}}^{\dagger}P_{\mathbf{k'}}P_{\mathbf{r}}\right)+\left(P_{\mathbf{i}}^{\dagger}P_{\mathbf{k}}^{10}\left|1\right\rangle \left\langle 0\right|P_{\mathbf{i}}\otimes P_{\mathbf{s}}^{\dagger}P_{\mathbf{k'}}P_{\mathbf{r}}\right)+\left(P_{\mathbf{i}}^{\dagger}P_{\mathbf{k}}^{01}\left|0\right\rangle \left\langle 1\right|P_{\mathbf{i}}\otimes P_{\mathbf{r}}^{\dagger}P_{\mathbf{k'}}P_{\mathbf{s}}\right)+\left(P_{\mathbf{i}}^{\dagger}P_{\mathbf{k}}^{11}\left|1\right\rangle \left\langle 1\right|P_{\mathbf{i}}\otimes P_{\mathbf{s}}^{\dagger}P_{\mathbf{k'}}P_{\mathbf{s}}\right)\right],
\end{equation}
\normalsize
where $P_{\mathbf{k}}^{ij}$ represents the $\left(i,j\right)$ matrix
element of $P_{\mathbf{k}}$. From Eq. (\ref{eq:paulisproofs}) it directly
follows that the first and fourth elements (corresponding to diagonal
terms of the control) vanish. Moreover, observe that any Pauli operator
$P_{\mathbf{k}}$ (see previous section) is represented by a matrix
which is either diagonal, or zero-diagonal (i.e. all diagonal elements
are 0). In the former case, the second and third elements directly
vanish. In the latter one, and noting that $\sum_{\mathbf{r,s}}P_{\mathbf{s}}^{\dagger}P_{\mathbf{k'}}P_{\mathbf{r}}=\sum_{\mathbf{r,s}}P_{\mathbf{r}}^{\dagger}P_{\mathbf{k'}}P_{\mathbf{s}}$,
the Pauli operator $P_{\mathbf{k}}$ is recovered from all its non-diagonal
components, such that
\small
\begin{equation}
\sum_{\mathbf{i,r,s}}\left[\left(P_{\mathbf{i}}^{\dagger}P_{\mathbf{k}}^{10}\left|1\right\rangle \left\langle 0\right|P_{\mathbf{i}}\otimes P_{\mathbf{s}}^{\dagger}P_{\mathbf{k'}}P_{\mathbf{r}}\right)+\left(P_{\mathbf{i}}^{\dagger}P_{\mathbf{k}}^{01}\left|0\right\rangle \left\langle 1\right|P_{\mathbf{i}}\otimes P_{\mathbf{r}}^{\dagger}P_{\mathbf{k'}}P_{\mathbf{s}}\right)\right]=\sum_{\mathbf{i}}P_{\mathbf{i}}^{\dagger}P_{\mathbf{k}}P_{\mathbf{i}}\otimes\sum_{\mathbf{p,q}}P_{\mathbf{p}}^{\dagger}P_{\mathbf{k'}}P_{\mathbf{q}}.
\end{equation}\normalsize
In this case, the left part of the tensor product (from the control
subspace) vanishes due to Eq. (\ref{eq:paulisproofs}), and condition
Eq. (\ref{eq: conditionAp}) is proven to be fulfilled for any $\mathbf{\left\{ k,k'\right\} }\in\mathbb{Z}_{d}^{2}$.
Observe that the reasoning of this proof can be generalized
for a qudit control register, as well as for a multi control scenario,
for an arbitrary dimension and number of control systems. In particular, for two control qubits, the Toffoli gate
is included among the set of gates of the form:
\small
\begin{equation}
T_{\mathbf{i,j,s,r,p,q}}=\left(P_{\mathbf{i}}\otimes P_{\mathbf{j}}\otimes I\right)\left(\left|0\right\rangle _{c_{1}}\left\langle 0\right|\otimes\left|0\right\rangle _{c_{2}}\left\langle 0\right|\otimes P_{\mathbf{s}}+\left|0\right\rangle _{c_{1}}\left\langle 0\right|\otimes\left|1\right\rangle _{c_{2}}\left\langle 1\right|\otimes P_{\mathbf{r}}+\left|1\right\rangle _{c_{1}}\left\langle 1\right|\otimes\left|0\right\rangle _{c_{2}}\left\langle 0\right|\otimes P_{\mathbf{p}}+\left|1\right\rangle _{c_{1}}\left\langle 1\right|\otimes\left|1\right\rangle _{c_{2}}\left\langle 1\right|\otimes P_{\mathbf{q}}\right).
\end{equation}
\normalsize

\subsubsection*{Multipartite Mølmer--Sørensen type gates}

Another instance of particular importance that can be benchmarked
with the coherent approach are multipartite Mølmer--Sørensen type
operations, which are a kind of multipartite entangling gates.
Concretely, given the set of operations
\begin{equation}
M_{\mathbf{i}}=P_{\mathbf{i}}U_{n}\left(\theta\right),\label{eq:set MS}
\end{equation}
for all the $P_{\mathbf{i}}$  Pauli elements of the form Eq.  (\ref{eq:paulielement})
with $\mathbf{i}\in\mathbb{Z}_{2}^{2n}$, it fulfills Eq. (\ref{eq: conditionAp}),
where $U_{n}\left(\theta\right)$ is the multipartite Mølmer--Sørensen gate \cite{MolmerSor}, i.e.
\begin{equation}
MS=U_{n}\left(\theta\right)=\prod_{s=1}^{n-1}\prod_{r=s+1}^{n}e^{i\theta X^{(s)}\otimes X^{(r)}},
\end{equation}
where $X^{(s)}$ defines the single-qubit Pauli $X$ gate acting on
qubit $s$. This fits for an arbitrary angle $\theta$, which defines
the entangling power of the operation. We can prove that Eq. (\ref{eq: conditionAp}),
i.e.
\begin{equation}
\sum_{\mathbf{i}}M_{\mathbf{i}}^{\dagger}P_{\mathbf{j}}M_{\mathbf{i}}=\begin{cases}
I_{d}^{\otimes n} & \mathbf{j}=\mathbf{o}\\
0 & \mathbf{j}\neq\mathbf{o}
\end{cases},\label{eq:condiitonapp}
\end{equation}
is fulfilled observing that
\begin{equation}
\sum_{\mathbf{i}}M_{\mathbf{i}}^{\dagger}P_{\mathbf{j}}M_{\mathbf{i}}=\sum_{\mathbf{i}}\left(P_{\mathbf{i}}U_{n}\left(\theta\right)\right)^{\dagger}P_{\mathbf{j}}\left(P_{\mathbf{i}}U_{n}\left(\theta\right)\right).\label{eq:MSinter}
\end{equation}
By expansion, we find
\begin{equation}
U_{n}\left(\theta\right)P_{\mathbf{j}}U_{n}\left(\theta\right)^{\dagger}=\prod_{s=1}^{n-1}\prod_{r=s+1}^{n}e^{-i\theta X^{(s)}\otimes X^{(r)}}P_{\mathbf{j}}\prod_{s'=1}^{n-1}\prod_{r'=s'+1}^{n}e^{i\theta X^{(s')}\otimes X^{(r')}}.\label{eq:MSPauli}
\end{equation}
Taking into account that
\begin{equation}
e^{i\theta P_{1}^{(1)}\otimes\cdots\otimes P_{n}^{(n)}}=\cos\left(\theta\right)I^{\otimes n}+i\sin\left(\theta\right)P_{1}^{(1)}\otimes\cdots\otimes P_{n}^{(n)},
\end{equation}
and the fact that $e^{i\theta P_{i}}e^{i\theta P_{j}}=e^{i\theta\left(P_{i}+P_{j}\right)}$
only if $\left[P_{i},P_{j}\right]=0$, it is therefore easy to see
that Eq. (\ref{eq:MSPauli}) maps any Pauli element $P_{\mathbf{j}}\neq P_{\mathbf{o}}$
into another (or a linear combination of) Pauli element $P_{\mathbf{j'}}\neq P_{\mathbf{o}},$
and $P_{\mathbf{o}}$ is always mapped to $P_{\mathbf{o}}$. Hence
Eq. (\ref{eq:MSinter}) is transformed into $\sum_{\mathbf{i}}P_{\mathbf{i}}^{\dagger}P_{\mathbf{j'}}P_{\mathbf{i}},$
and given the property Eq. (\ref{eq:paulisproofs}), it is direct that
Eq. (\ref{eq:condiitonapp}) is fulfilled. Note that same arguments
are valid for variations of the MS gate, substituting $XX$ interactions
by e.g. $ZZ$ or $YY$ interactions.

\setcounter{equation}{0}
\renewcommand\theequation{C.\arabic{equation}}

\subsubsection*{Arbitrary $n-$qudit unitary sets}

The previous setting with the MS gate can be generalized for any arbitrary (set of) $n-$qudit unitary operation(s). Consider the set of operations given by the elements
\begin{equation}
M_{\mathbf{i}}=P_{\mathbf{i}}U,\label{eq:set anyU}
\end{equation}
for any $n-$qudit unitary $U$ and the set of all $n-$qudit Pauli operators $P_{\mathbf{i}} \in \mathcal{P}_{d,n}$. Given the fact that the trace of an operator is invariant under conjugation by another unitary and given Eq. (\ref{eq:paulisproofs}), it can be easily seen that this set fulfills Eq. (\ref{eq: conditionAp}), i.e.
\begin{equation}
\sum_{\mathbf{i}}M_{\mathbf{i}}^{\dagger}P_{\mathbf{j}}M_{\mathbf{i}}=\begin{cases}
I_{d}^{\otimes n} & \mathbf{j}=\mathbf{o}\\
0 & \mathbf{j}\neq\mathbf{o}
\end{cases}.\label{eq:condiitonapp2}
\end{equation}
We can therefore benchmark any unitary (set of) operation(s) followed by some Pauli rotation. In particular, observe that, under the assumption that Pauli operations are implemented perfectly, one can directly access to the (average) error of the arbitrary unitary gate(s). This assumption can be justified from the perspective that multi-qudit Pauli rotations are experimentally much easier and more reliable to implement than arbitrary unitaries as e.g. an entangling gate like the MS gate or a multi-qudit Clifford gate.

\subsection*{Appendix C: Averaged coherent gate fidelity derivation}

The gate fidelity between two quantum operations $\mathcal{U},\mathcal{E}$
is defined as \cite{Magesan11_2}

\begin{equation}
F_{\mathcal{U},\mathcal{E}}=\mathrm{tr}\left(\mathcal{U}\left(\left|\varphi\right\rangle \left\langle \varphi\right|\right)\mathcal{E}\left(\left|\varphi\right\rangle \left\langle \varphi\right|\right)\right),
\end{equation}
for some state $\left|\varphi\right\rangle \left\langle \varphi\right|.$
By setting $\Lambda=\mathcal{U}^{\dagger}\circ\mathcal{E}$, one can
define the average gate fidelity of the noisy channel $\varLambda$, which
characterize the fidelity of the ideal quantum gate with respect to
its imperfect implementation, i.e.
\begin{equation}
\bar{F}_{\Lambda,\mathcal{I}}=\int d\varphi\mathrm{tr}\left(\left|\varphi\right\rangle \left\langle \varphi\right|\Lambda\left(\left|\varphi\right\rangle \left\langle \varphi\right|\right)\right),
\end{equation}
with the average over the invariant Haar measure on pure states. In particular, given
e.g. the Pauli decomposition of a general noisy channel of Eq. (\ref{eq:cptp-1}),
it can be shown \cite{Magesan11_2} that
\begin{equation}
\bar{F}_{\Lambda,\mathcal{I}}=\frac{d\chi_{00}+1}{d+1},
\end{equation}
where $\chi_{00}$ is the $\left(00\right)$ element of the Pauli $\chi-$matrix
that we use to describe the noise.

We show here how the expression for the average  sequence coherent fidelity, which directly relates to $\chi_{00}$ in the zeroth order approximation, is found given a set of operations that
fulfills Eq. (\ref{eq: conditionAp}). Consider a set of operations
$G$ of size $\text{|}G|$. Consider now the general definition for
the average sequence coherent  fidelity of a sequence of length $m$
given by
\begin{equation}
F_{G}\left(m,|G|^{m}\right)=\mathrm{tr}\left(E_{\psi}\rho_{f}\right),\label{eq:fideap1}
\end{equation}
where the number of elements in
the superposition (also the size of the control register) is $\text{|}G|^{m}$,
comprising \textit{all} the possible different sequences of length
$m$. Let us assume for the moment that the control register has no
error associated, i.e. ideal control implementation. Therefore:

\begin{equation}
F_{G}\left(m,|G|^{m}\right)=\mathrm{tr}\left(E_{\psi}\left(\hat{I}\otimes {\xi}^{(m+1)}\right)\circ\hat{CU}^{(m+1)}\circ\left(\bigcirc_{r=1}^{m}\left[ {\xi}\circ\hat{CU}^{(r)}\right]\left(\rho_{0}\right)\right)\right),\label{eq:fideap2}
\end{equation}
where $\xi$ is
a general error map, and the POVM
corresponds to $\left\{ E_{\psi},I-E_{\psi}\right\} \approx\left\{ \left|\psi\right\rangle \left\langle \psi\right|,I-\left|\psi\right\rangle \left\langle \psi\right|\right\} $ taking into account measurement imperfections,
with $\left|\psi\right\rangle =\left|+\right\rangle _{c}^{|G|^{m}}\otimes\left|\varphi\right\rangle _{in}$ and where the initial state reads

\begin{equation}
\rho_{0}=\frac{1}{k}\sum_{i,j=0}^{k-1}\left|i\right\rangle _{c}\left\langle j\right|\otimes\left|\varphi\right\rangle _{in}\left\langle \varphi\right|,\label{eq:initialstate}
\end{equation}
where the control register is initialized in the state $\left|+\right\rangle _{c}^{k}=\frac{1}{\sqrt{k}}\sum_{i=0}^{k-1}\left|i\right\rangle _{c}$
for some $k$, and where $\left|\varphi\right\rangle _{in}\left\langle \varphi\right|\approx\left|0\right\rangle \left\langle 0\right|$ taking into account state preparation errors. We recall that we restrict
to the case of position-independent gate noise, i.e. zeroth-order approximation.  The contribution of the
diagonal elements with respect to the register state are the equivalent
to the standard classical average over \textit{all} the possible sequences, while the
contribution of coherent terms leads to an extra information gain. We can rewrite Eq. (\ref{eq:fideap2}) as:

\begin{equation}
F_{G}\left(m,|G|^{m}\right)=\frac{1}{|G|^{m}}\sum_{i,j=1}^{|G|^{m}}\mathrm{tr}\left(E_{\psi}\left(\hat{I}\otimes {\xi}^{(m+1)}\right)\circ\hat{CU}^{(m+1)}\circ\left(\bigcirc_{r=1}^{m}\left[ {\xi}\circ\hat{CU}^{(r)}\right]\left(\left|i\right\rangle _{c}\left\langle j\right|\otimes\left|\varphi\right\rangle _{in}\left\langle \varphi\right|\right)\right)\right).\label{eq:fideap3}
\end{equation}
Observe that we equivalently define here the sum over $i,j$ from
$1$ to $|G|^{m}$ for simplicity. The fidelity reads:

\begin{equation}
F_{G}\left(m,|G|^{m}\right)=\frac{1}{|G|^{m}}\sum_{i,j=1}^{|G|^{m}}\mathrm{tr}\left(E_{\psi}\left(\hat{I}\otimes {\xi}^{(m+1)}\right)\left(U_{i}^{(m)}\cdots U_{i}^{(1)}\right)^{\dagger}\left\{ \prod_{r=1}^{m} {\xi}\circ\left[U_{i}^{(r)}\left(\left|i\right\rangle _{c}\left\langle j\right|\otimes\left|\varphi\right\rangle _{in}\left\langle \varphi\right|\right)U_{j}^{(r)^{\dagger}}\right]\right\} \left(U_{j}^{(m)}\cdots U_{j}^{(1)}\right)\right),\label{eq:fideap4}
\end{equation}
where all unitaries $U_{s}^{(r)}\in G$ and the error act onto the
main register. Each unitary $U_{s}^{(r)}$ with $s=\left\{ 0,\ldots,|G|^{m}-1\right\} $
and $r=\left\{ 1,\ldots,m\right\} $ is appropriately chosen in such
a way that each sequence $U_{s}^{(1)}\cdots U_{s}^{(m)}$ corresponds
to each one of all the $|G|^{m}$ different possible sequences of
length $m.$ We can expand the $r=m^{th}$  position as:
\small
\begin{multline}
F_{G}\left(m,|G|^{m}\right)= \\
\frac{1}{|G|^{2m}}\sum_{i,j=1}^{|G|^{m-1}}\mathrm{tr}\left(E_{\psi} {\xi}^{(m+1)}\left(U_{i}^{(m-1)}\cdots U_{i}^{(1)}\right)^{\dagger}\left\{ \sum_{s,q=1}^{|G|}\sum_{l,l'=1}^{d^{2n}}\chi_{l,l'}U_{s}^{(m)^{\dagger}}\sigma_{l}U_{s}^{(m)}\left[\rho_{ij}^{(m-1)}\right]U_{q}^{(m)^{\dagger}}\sigma_{l'}^{\dagger}U_{q}^{(m)}\right\} \left(U_{j}^{(m-1)}\cdots U_{j}^{(1)}\right)\right),
\label{eq:fidelfinal1}
\end{multline}\normalsize
with $\rho_{ij}^{(m-1)}=\prod_{r=1}^{m-1} {\xi}\circ\left[U_{i}^{(r)}\left(\left|\varphi\right\rangle _{in}\left\langle \varphi\right|\right)U_{j}^{(r)^{\dagger}}\right]$,
and the factor $\frac{1}{|G|^{m}}$ coming from the projective measurement
corresponding of the control register subspace. This expression can be understood as follows. For a
sequence length $m$, one can find $|G|^{2m}$ different components
of the density matrix $\rho_{f}^{(m)}$, with $|G|^{m}$ diagonal
elements, corresponding to all the possible classical-mixture sequences, and
$\left(|G|^{2m}-|G|^{m}\right)$ coherent elements. Note now that
for sequence length $m-1$, the density matrix $\rho_{ij}^{(m-1)}$
has $|G|^{2m-2}$ different elements, corresponding to the sum over
$i,j$ from $1$ to $|G|^{m-1}$. Therefore, the sum expansion of
Eq. (\ref{eq:fidelfinal1}) is appropriately justified.

The Pauli operators of the noise map are given by $\sigma_{\mathbf{x}}\equiv P_{\mathbf{x}}\equiv X^{\mathbf{x}_{i}}Z^{\mathbf{x}_{j}}$
(see Appendix A) for $n$ number of qudits.
Consider now
\begin{equation}
\sum_{i=1}^{|G|}U_{i}^{\dagger}\sigma_{\mathbf{j}}U_{i}=\begin{cases}
|G|I_{d}^{\otimes n} & \mathbf{j}=\mathbf{o}\\
0 & \mathbf{j}\neq\mathbf{o}
\end{cases},\label{eq: condition-1}
\end{equation}
where $\sigma_{\mathbf{o}}=I^{\otimes n}$. If a gate set can be benchmarked via coherent RB, then this condition is fulfilled. We can apply the decomposition of Eq. (\ref{eq:fidelfinal1}) recursively
for each $r=m,\ldots,1$. For each $r$ only survives the term associated
to $\chi_{00}$ with a factor $|G|^{2}.$ Hence it is direct to see
that the resulting fidelity reads
\begin{equation}
F_{G}\left(m,|G|^{m}\right)=A\chi_{00}^{m},\label{eq:finalfidelity-1}
\end{equation}
with $A=\mathrm{tr}\left(E_{\psi} {\xi}^{(m+1)}(\left|\varphi\right\rangle _{in}\left\langle \varphi\right|)\right).$

\subsubsection*{Example}
For a better understanding, we provide a simple example. Consider the set $G$ consisting of two unitary gates $U_{1},U_{2}$
and a sequence length $m=2.$ Ignoring preparation errors, the initial
state reads
\begin{equation}
\rho_{0}=\frac{1}{4}\sum_{i,j=0}^{2^{2}-1}\left|i\right\rangle _{c}\left\langle j\right|\otimes\left|0\right\rangle \left\langle 0\right|,
\end{equation}
where the single-qubit target  is simply prepared in the  pure
state $\rho_{in}=\left|0\right\rangle \left\langle 0\right|$, and
the control in the $2^{2}-$dimensional $\left|+\right\rangle $ state,
since $\text{|}G|=2$ is the size of the set, and we consider all
the possible sequences of length $m$, i.e. $\text{|}G|^{m}=2^{2}$.
 The next
step consists in applying controlled operations of the form $
CU^{(j)}=\sum_{i=0}^{2^{2}-1}\left|i\right\rangle _{c}\left\langle i\right|\otimes U_{i}^{(j)}$,
where $j=\left\{ 1,2\right\}$ defines the position, noting that
for different values of $j$, the controlled operations changes, such
that all possible combinations are invoked. The resulting state assuming
no noise for the moment reads
\begin{equation}
\rho=CU^{(2)}CU^{(1)}\rho_{0}CU^{\dagger(1)}CU^{\dagger(2)}=\frac{1}{4}\sum_{i,j=0}^{2^{2}-1}\left[\left|i\right\rangle _{c}\left\langle j\right|\otimes U_{i}^{(2)}U_{i}^{(1)}\left|0\right\rangle \left\langle 0\right|U_{j}^{\dagger(1)}U_{j}^{\dagger(2)}\right].\label{eq:eqq}
\end{equation}
Observe again that $U_{r}^{(s)}=\left\{ U_{1},U_{2}\right\} $ is
not necessarily equal to $U_{r}^{(s')}$ if $s\neq s'$. We can also
equivalently relabel the last expression in a more clear notation, i.e.
\begin{equation}
\rho=\frac{1}{4}\sum_{q_{1},q_{1}',q_{2},q'_{2}=1}^{2}\left[\left|q_{1},q_{2}\right\rangle _{c}\left\langle q'_{1},q'_{2}\right|\otimes U_{q_{2}}U_{q_{1}}\left|0\right\rangle \left\langle 0\right|U_{q'_{1}}^{\dagger}U_{q'_{2}}^{\dagger}\right].\label{eq:eq2}
\end{equation}
Note also that, although $\text{|}G|^{m}=2^{2}=4$ different sequences
are considered, we actually find $\text{|}G|^{2m}=16$ terms in Eq.
(\ref{eq:eq2}), including the coherent terms. As a function of gates
$U_{1},U_{2}$ we can rewrite Eq. (\ref{eq:eqq}) and Eq. (\ref{eq:eq2}) as
\small
\begin{equation}
\rho=\frac{1}{4}\left(\begin{array}{cccc}
\left|0\right\rangle _{c}\left\langle 0\right|\otimes U_{1}U_{1}\rho_{in}U_{1}^{\dagger}U_{1}^{\dagger} & \left|0\right\rangle _{c}\left\langle 1\right|\otimes U_{1}U_{1}\rho_{in}U_{1}^{\dagger}U_{2}^{\dagger} & \left|0\right\rangle _{c}\left\langle 2\right|\otimes U_{1}U_{1}\rho_{in}U_{2}^{\dagger}U_{1}^{\dagger} & \left|0\right\rangle _{c}\left\langle 3\right|\otimes U_{1}U_{1}\rho_{in}U_{2}^{\dagger}U_{2}^{\dagger}\\
\left|1\right\rangle _{c}\left\langle 0\right|\otimes U_{2}U_{1}\rho_{in}U_{1}^{\dagger}U_{1}^{\dagger} & \left|1\right\rangle _{c}\left\langle 1\right|\otimes U_{2}U_{1}\rho_{in}U_{1}^{\dagger}U_{2}^{\dagger} & \left|1\right\rangle _{c}\left\langle 2\right|\otimes U_{2}U_{1}\rho_{in}U_{2}^{\dagger}U_{1}^{\dagger} & \left|1\right\rangle _{c}\left\langle 3\right|\otimes U_{2}U_{1}\rho_{in}U_{2}^{\dagger}U_{2}^{\dagger}\\
\left|2\right\rangle _{c}\left\langle 0\right|\otimes U_{1}U_{2}\rho_{in}U_{1}^{\dagger}U_{1}^{\dagger} & \left|2\right\rangle _{c}\left\langle 1\right|\otimes U_{1}U_{2}\rho_{in}U_{1}^{\dagger}U_{2}^{\dagger} & \left|2\right\rangle _{c}\left\langle 2\right|\otimes U_{1}U_{2}\rho_{in}U_{2}^{\dagger}U_{1}^{\dagger} & \left|2\right\rangle _{c}\left\langle 3\right|\otimes U_{1}U_{2}\rho_{in}U_{2}^{\dagger}U_{2}^{\dagger}\\
\left|3\right\rangle _{c}\left\langle 0\right|\otimes U_{2}U_{2}\rho_{in}U_{1}^{\dagger}U_{1}^{\dagger} & \left|3\right\rangle _{c}\left\langle 1\right|\otimes U_{2}U_{2}\rho_{in}U_{1}^{\dagger}U_{2}^{\dagger} & \left|3\right\rangle _{c}\left\langle 2\right|\otimes U_{2}U_{2}\rho_{in}U_{2}^{\dagger}U_{1}^{\dagger} & \left|3\right\rangle _{c}\left\langle 3\right|\otimes U_{2}U_{2}\rho_{in}U_{2}^{\dagger}U_{2}^{\dagger}
\end{array}\right).\label{eq: matrix}
\end{equation}
\normalsize
Observe that the diagonal terms correspond to the four possible combinations
of $\hat{U}_{i}\circ\hat{U}_{j}$ with $i,j=\left\{ 1,2\right\} .$
These incoherent elements correspond to the classical mixture of sequences
considered in the standard approach. The additional (non-diagonal) coherent terms do not appear in the standard approach, and we make use of them in the following.
Finally, a controlled operation
is applied invoking the inverses of each sequence, i.e. in our case:
\begin{equation}
CU^{(m+1)}=\left|0\right\rangle _{c}\left\langle 0\right|\otimes\left(U_{1}U_{1}\right)^{\dagger}+\left|1\right\rangle _{c}\left\langle 1\right|\otimes\left(U_{1}U_{2}\right)^{\dagger}+\left|2\right\rangle _{c}\left\langle 2\right|\otimes\left(U_{2}U_{1}\right)^{\dagger}+\left|3\right\rangle _{c}\left\langle 3\right|\otimes\left(U_{2}U_{2}\right)^{\dagger},
\end{equation}
such that every term $\left(i,j\right)$ in Eq. (\ref{eq: matrix})
is mapped again to $\left|i\right\rangle _{c}\left\langle j\right|\otimes\left|0\right\rangle \left\langle 0\right|$
and the final state reads $\rho_{f}=\rho_{0}$.

Consider now the noisy case, i.e. after each $CU$ operation, a single-qubit
noise map affects the target register. We assume this noise map to
be independent of the position and the gate applied. We can write
the map in the Pauli basis as
\begin{equation}
\xi\left(\rho\right)=\sum_{l,l'=0}^{3}\chi_{l,l'}\sigma_{l}\rho\sigma_{l'},
\end{equation}
which represents a completely general CPTP single-qubit channel. Therefore Eq. (\ref{eq:eqq}) becomes
\begin{equation}
\rho=\frac{1}{4}\sum_{i,j=0}^{3}\sum_{l_{1},l_{1}',l_{2},l'_{2}=0}^{3}\chi_{l_{1},l'_{1}}\chi_{l_{2},l'_{2}}\left[\left|i\right\rangle _{c}\left\langle j\right|\otimes\sigma_{l_{2}}U_{i}^{(2)}\sigma_{l_{1}}U_{i}^{(1)}\left|0\right\rangle \left\langle 0\right|U_{j}^{\dagger(1)}\sigma_{l'_{1}}U_{j}^{\dagger(2)}\sigma_{l'_{2}}\right].\label{eq:eqq-1}
\end{equation}
After the $CU^{(m+1)}$ operation, that we assume here to be noiseless
for simplicity:
\small
\begin{equation}
\rho_{f}=\frac{1}{4}\sum_{i,j=0}^{3}\sum_{l_{1},l_{1}',l_{2},l'_{2}=0}^{3}\chi_{l_{1},l'_{1}}\chi_{l_{2},l'_{2}}\left[\left|i\right\rangle _{c}\left\langle j\right|\otimes\left(U_{i}^{(2)}U_{i}^{(1)}\right)^{\dagger}\sigma_{l_{2}}U_{i}^{(2)}\sigma_{l_{1}}U_{i}^{(1)}\left|0\right\rangle \left\langle 0\right|U_{j}^{\dagger(1)}\sigma_{l'_{1}}U_{j}^{\dagger(2)}\sigma_{l'_{2}}\left(U_{j}^{(2)}U_{j}^{(1)}\right)\right].\label{eq:mid0}
\end{equation} \normalsize
A final projective measurement $\left\{ E_{\psi},I-E_{\psi}\right\} =\left\{ \left|\psi\right\rangle \left\langle \psi\right|,I-\left|\psi\right\rangle \left\langle \psi\right|\right\} $,
with $\left|\psi\right\rangle =\left|+\right\rangle _{c}^{d=4}\otimes\left|0\right\rangle $
is performed. Therefore
\small
\begin{equation}
F(\rho_{f})=\frac{1}{16}\sum_{i,j=0}^{2^{2}-1}\sum_{l_{1},l_{1}',l_{2},l'_{2}=0}^{3}\chi_{l_{1},l'_{1}}\chi_{l_{2},l'_{2}}\mathrm{tr}\left(E_{0}\left[\left(U_{i}^{(2)}U_{i}^{(1)}\right)^{\dagger}\sigma_{l_{2}}U_{i}^{(2)}\sigma_{l_{1}}U_{i}^{(1)}\left|0\right\rangle \left\langle 0\right|U_{j}^{\dagger(1)}\sigma_{l'_{1}}U_{j}^{\dagger(2)}\sigma_{l'_{2}}\left(U_{j}^{(2)}U_{j}^{(1)}\right)\right]\right),\label{eq:mid}
\end{equation} \normalsize
where a factor $\frac{1}{\text{|}G|^{m}}=\frac{1}{4}$ comes from
the measurement of the control register. As before, given the fact
that the control register has been measured out, we can simply relabel
the expression in a more clear notation as
\small
\begin{equation}
F(\rho_{f})=\frac{1}{16}\sum_{q_{1},q_{1}',q_{2},q'_{2}=1}^{2}\sum_{l_{1},l_{1}',l_{2},l'_{2}=0}^{3}\chi_{l_{1},l'_{1}}\chi_{l_{2},l'_{2}}\mathrm{tr}\left(E_{0}\left[\left(U_{q_{2}}U_{q_{1}}\right)^{\dagger}\sigma_{l_{2}}U_{q_{2}}\sigma_{l_{1}}U_{q_{1}}\left|0\right\rangle \left\langle 0\right|U_{q'_{1}}^{\dagger}\sigma_{l'_{1}}U_{q'_{2}}^{\dagger}\sigma_{l'_{2}}\left(U_{q'_{2}}U_{q'_{1}}\right)\right]\right).
\end{equation} \normalsize
All the elements corresponding to the first sequence position can
be compacted for notation simplicity such that
\begin{equation}
F(\rho_{f})=\frac{1}{16}\sum_{q_{1},q_{1}',q_{2},q'_{2}=1}^{2}\sum_{l_{2},l'_{2}=0}^{3}\chi_{l_{2},l'_{2}}\mathrm{tr}\left(E_{0}\left[\left(U_{q_{2}}U_{q_{1}}\right)^{\dagger}\sigma_{l_{2}}U_{q_{2}}\rho_{q_{1},q'_{1}}U_{q'_{2}}^{\dagger}\sigma_{l'_{2}}\left(U_{q'_{2}}U_{q'_{1}}\right)\right]\right).
\end{equation}
Observe now that, if condition
\begin{equation}
\sum_{i=1}^{2}U_{i}^{\dagger}\sigma_{j}U_{i}=\begin{cases}
|G|I & j=0\\
0 & j\neq0
\end{cases},\label{eq: condition-2}
\end{equation}
where $\sigma_{0}=I$, is fulfilled, only the $\chi_{l_{2},l'_{2}}=\chi_{00}$
term survives, with an overall factor $|G|^{2}=4.$ By repeating the
same arguments for the remaining position $j=1$, we end up with:
\begin{equation}
F(\rho_{f})=\frac{1}{16}|G|^{4}\chi_{00}^{2}=\chi_{00}^{2}.
\end{equation}

\setcounter{equation}{0}
\renewcommand\theequation{D.\arabic{equation}}
\subsection*{Appendix D: Coherent IRB. Benchmark a $n-$qudit Clifford gate with Paulis}
Interleaved randomized benchmarking (IRB) consists in interleaving a particular gate every second position of the sequence, and comparing with respect to a reference sequence such that the noise parameter of the particular gate can be extracted. We show here a detailed description of the process of  benchmarking a $n-$qudit Clifford gate using the set of Pauli operators. The reference process of the IRB protocol proceeds as the RB process of Appendix C with the $n-$qudit Pauli set. The interleaved process proceeds as follows. Every second position, the $C$ gate is applied in all
the branches of the superposition. We denote this operation as
$CU_{C}=I\otimes C=U_{C}.$  The remaining controlled operations are again of the form $CU^{(j)}=\sum_{i=0}^{k-1}\left|i\right\rangle _{c}\left\langle i\right|\otimes U_{i}^{(j)},$
with $U_{i}^{(j)}\subseteq \mathcal{P}_{d,n}.$ For a known initial state $\rho_{0}=\left|+\right\rangle _{c}^{d}\left\langle +\right|\otimes\left|\varphi\right\rangle _{in}\left\langle \varphi\right|$, the final state reads
\begin{equation}
\rho_{f}=\hat{CU}^{(m+1)}\circ\left(\bigcirc_{j=1}^{m}\left[ {\xi_{C}}\circ\hat{U}_{C}\circ {\xi}\circ\hat{CU}^{(j)}\right]\left(\rho_{0}\right)\right).\label{eq:interv2b}
\end{equation}
Following a similar reasoning than in the RB process, we write the average sequence fidelity decomposing the $m^{th}$
sequence position considering \textit{all} possible sequences in superposition:
\small
\begin{multline}
F_{P,C}\left(m,|P|^{m}\right)=\frac{1}{|P|^{2m}} \sum_{i,j=1}^{|P|^{m-1}} \\
\mathrm{tr}\left(E_{\psi}\left(U_{i}^{(m-1)}C\cdots U_{i}^{(1)}C\right)^{\dagger}\left\{ \sum_{s,q=1}^{|P|}\sum_{l_{1},l'_{1},l_{2},l'_{2}=1}^{d^{2n}}\chi_{l_{1},l_{1}'}^{C}\chi_{l_{2},l_{2}'}^{P}U_{s}^{(m)^{\dagger}}C^{\dagger} P_{l_{1}}C P_{l_{2}}U_{s}^{(m)}\left[\rho_{ij}^{(m-1)}\right]U_{q}^{(m)^{\dagger}} P_{l'_{2}}C^{\dagger} P_{l'_{1}}CU_{q}^{(m)}\right\} \left(U_{i}^{(m-1)}C\cdots U_{i}^{(1)}C\right)\right),\label{eq:Fpc}
\end{multline}
\normalsize
where we have measured the control register, and note that in our
case $|P|=d^{2n}$ too.  Note also that, due to the controlled nature of the
setting, one cannot analyze the evolution of the main register as
a function of concatenated quantum maps. $\chi^{P}$ and $\chi^{C}$
are the corresponding noisy channel matrices of the Pauli set and
the $C$ gate respectively. From step one, we already know $\chi_{00}^{P}=\chi_{00}^{ref}$
and $\chi_{00}^{C}$ is the objective parameter that the protocol
aims to find. From Eq. (\ref{eq:Fpc}), observe that, by definition
of Clifford operation, $C^{\dagger} P_{i}C$
is mapped to an element of $\mathcal{P}\setminus I^{\otimes n}$ for $i \neq 0$
and to $I^{\otimes n}$ for $P_{0}=I^{\otimes n}.$ Therefore, the corresponding noise
matrix is mapped to $\chi^{C}\rightarrow\chi^{\bar{C}}$, noting that
$\chi_{00}^{C}=\chi_{00}^{\bar{C}}$. The resulting state reads
\small
\begin{multline}
F_{P,C}\left(m,|P|^{m}\right)=\frac{1}{|P|^{2m}} \sum_{i,j=1}^{|P|^{m-1}} \\
\mathrm{tr}\left(E_{\psi}\left(U_{i}^{(m-1)}C\cdots U_{i}^{(1)}C\right)^{\dagger}\left\{ \sum_{s,q=1}^{|P|}\sum_{l_{1},l'_{1},l_{2},l'_{2}=1}^{d^{2n}}\chi_{l_{1},l_{1}'}^{\bar{C}}\chi_{l_{2},l_{2}'}^{P}U_{s}^{(m)^{\dagger}} P_{l_{1}} P_{l_{2}}U_{s}^{(m)}\left[\rho_{ij}^{(m-1)}\right]U_{q}^{(m)^{\dagger}} P_{l'_{2}} P_{l'_{1}}U_{q}^{(m)}\right\} \left(U_{i}^{(m-1)}C\cdots U_{i}^{(1)}C\right)\right).\label{eq:Fpc2}
\end{multline}
\normalsize
Finally, following exactly the same reasoning that for the coherent RB, we find that the fidelity follows a decay curve
of the form
\begin{equation}
F_{P,C}\left(m\right)=\left(\chi_{00}^{P\circ\bar{C}}\right)^{m},\label{eq:Fpc3b}
\end{equation}
where $\chi_{00}^{P\circ\bar{C}}=\sum_{i,j}\chi_{ij}^{P}\chi_{ij}^{\bar{C}}$.

One can obtain an estimation of the desired
parameter $\chi_{00}^{C}$ noting that $\chi_{00}^{C}=\chi_{00}^{\bar{C}}$
and that (see \cite{Kimmel2014}):
\begin{equation}
\chi_{00}^{P\circ\bar{C}}=\chi_{00}^{P}\chi_{00}^{\bar{C}}\pm
\left[2\sqrt{\left(1-\chi_{00}^{P}\right)\chi_{00}^{P}\left(1-\chi_{00}^{\bar{C}}\right)\chi_{00}^{\bar{C}}}+\left(1-\chi_{00}^{P}\right)\left(1-\chi_{00}^{\bar{C}}\right)\right].\label{eq:Fpcbound}
\end{equation}
with some estimation bound which is tight
in the regime of interest, i.e. where Pauli gates fidelity is much larger than the
$C$ one. Note also that this bound is analogous in size to standard IRB approaches.

\setcounter{equation}{0}
\renewcommand\theequation{E.\arabic{equation}}

\subsection*{Appendix E: Statistical analysis and efficiency}
In a realistic setting considering a superposition of
all $|G|^{m}$ possible sequences is certainly inefficient. Therefore, the protocol is restricted to $k\ll|G|^{m}$ sequences, leading to an estimation of the average sequence fidelity with some deviation with respect to the exact value. Many works have studied the statistical properties of the standard RB in detail (see e.g. \cite{Wallman2014,Granade15,Harper19}). In particular, several bounds for the variance of
the process are found in different regimes, showing that a number
of around $k\approx 100$ sequences suffices in general for
good fidelity estimations of the Clifford group. In our case, this overhead arises in the dimension of the control register. Instead of performing
different sequential rounds and averaging over their fidelities, we
perform a single round step with the sequences in a coherent superposition. Notice that this corresponds to a reduction of the number of required experiments --and hence the required measurements to be performed-- by a factor that is given by the number $k$ of considered sequences in superposition.
Despite this reduction, we show indications that the estimation confidence in the superposed case can be even better in certain regimes of interest. Note however that one of the main advantages of the coherent RB is its much larger applicability, and therefore we can only restrict to the Clifford group for comparisons.

When considering a certain number $k$ of sequences in superposition which corresponds to the size of the control register, a confidence region of size $\epsilon$
arises, such that the probability that the estimated fidelity lies
within this confidence region is greater than some set confidence
level $1-\delta$, i.e.
\begin{equation}
P\left[\left|F_{k}-\bar{F}\right|<\epsilon\right]\geq1-\delta,\label{eq: interval}
\end{equation}
where $F_{k}$ is the estimated average fidelity given $k$ sequences
and $\bar{F}$ the exact averaged fidelity given by Eq. (\ref{eq:finalfidelity-1}).
We denote the difference $\left|F_{k}-\bar{F}\right|$ as the estimation
deviation. Note that expression Eq. (\ref{eq: interval}) also applies
for the standard case. In particular, in the standard approach, parameters
$\epsilon,\delta$ can be directly related to the number of sequences
$k$ and the variance $\sigma_{k}$ via concentration inequalities
(e.g. Hoeffding inequality). Such direct relations seem more complicated
to be found in the coherent case. Therefore, and also due to the heteroscedastic
behavior of the data in different regimes, we leave a detailed statistical
analysis for further work.

We provide a numerical analysis of the fidelity deviation (difference between the estimated numerical fidelity for few sequences and the analytical fidelity when considering all possible sequences) for different settings (see Fig. \ref{fig:deviation}).  We observe that,
in regimes of practical importance \cite{Wallman2014}, the upper bound of the fidelity deviation is always lower in the coherent approach. Moreover, we show that this fidelity deviation is comparable in magnitude if one goes out of the Clifford group to e.g. the Pauli group (where standard RB cannot be used to benchmark gates).

Although coherent RB outperforms standard RB in any regime, we find that a heuristic combination of standard and coherent RB analytical decay curves can even improve the results. For a given sequence length $m$, the number of possible sequences scales as $|G|^{m}$, where $|G|$ is the size of the gate-set. Therefore, in regimes where $k \ll |G|^{m}$, the coherence of the estimated fidelity is always much lower than in the analytical case. If the analytical decay curve is corrected taking into account the proximity to the incoherent case, i.e. $\bar{F}=\left(1-\frac{1}{k}\right) \bar{F}_{coherent}+ \frac{1}{k} \bar{F}_{standard}$, the performance can be enhanced (see Fig. \ref{fig:deviation} d).). Note however that this is only applicable to the Clifford case, since standard RB is not applicable for other groups or sets of gates.

Similar results are found for other noisy channels and different infidelity regimes.

\begin{figure}
\includegraphics[scale=0.5]{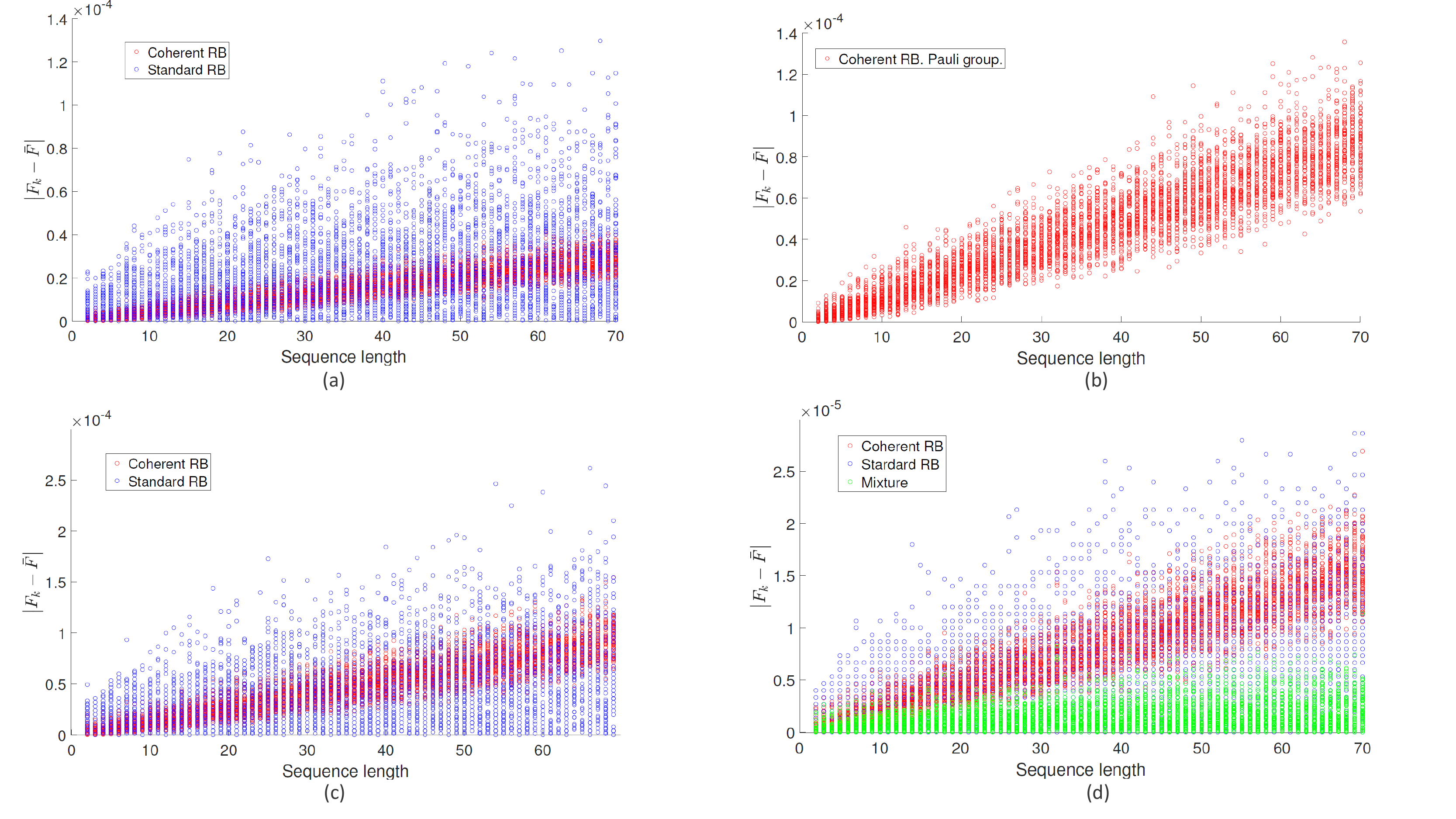}
\caption{ \label{fig:deviation} Numerical results of the estimation deviation
$\left|F_{k}-\bar{F}\right|$ as a function of the sequence length under dephasing noise for different gate sets and values of $k$.  For each sequence length, the protocol is performed $75$ independent times. a) Single-qubit Clifford group with $k=80$ and gate infidelity $10^{-4}$. Each point for coherent RB represents the single-run fidelity of a superposition of $80$ random sequences, whereas each point for standard RB represents the average over $80$ different random sequences.  b) Same parameters than the previous case but the group benchmarked is the single-qubit Pauli group. Note that fidelity deviation is comparable in magnitude to the Clifford case.  c) Clifford group with same parameter except $k=25$. The upper limit on the deviation for the standard case remains significantly larger than for the coherent case. d) Clifford case with gate infidelity of $10^{-5}$ and with $k=15$. Although coherent RB outperforms standard RB, results are improved considering a combination of both fitting curves.}
\end{figure}

\end{document}